\documentclass{aa}  

\usepackage{graphicx}
\usepackage{txfonts}
\usepackage{natbib,twoopt}
\usepackage[breaklinks=true]{hyperref} 
\bibpunct{(}{)}{;}{a}{}{,}             
\makeatletter
  \newcommandtwoopt{\citeads}[3][][]{\href{http://adsabs.harvard.edu/abs/#3}%
    {\def\hyper@linkstart##1##2{}%
     \let\hyper@linkend\@empty\citealp[#1][#2]{#3}}}
  \newcommandtwoopt{\citepads}[3][][]{\href{http://adsabs.harvard.edu/abs/#3}%
    {\def\hyper@linkstart##1##2{}%
     \let\hyper@linkend\@empty\citep[#1][#2]{#3}}}
  \newcommandtwoopt{\citetads}[3][][]{\href{http://adsabs.harvard.edu/abs/#3}%
    {\def\hyper@linkstart##1##2{}%
     \let\hyper@linkend\@empty\citet[#1][#2]{#3}}}
  \newcommandtwoopt{\citeyearads}[3][][]%
    {\href{http://adsabs.harvard.edu/abs/#3}
    {\def\hyper@linkstart##1##2{}%
     \let\hyper@linkend\@empty\citeyear[#1][#2]{#3}}}
\makeatother
\begin{document}

   \title{New nearby white dwarfs from Gaia DR1 TGAS and UCAC5/URAT\thanks{Partly based on observations with the 2.2m telescope of the German-Spanish Astronomical Centre at Calar Alto, Spain}}

   \subtitle{}

   \author{R.-D. Scholz
          \inst{1}
          \and
          H. Meusinger\inst{2,3}
          \and
          H. Jahrei{\ss}\inst{4}
          }

   \institute{Leibniz-Institut f\"ur Astrophysik Potsdam, 
              An der Sternwarte 16, D--14482 Potsdam, Germany\\
              \email{rdscholz@aip.de}
         \and
             Th\"uringer Landessternwarte Tautenburg, 
             Sternwarte 5, D--07778 Tautenburg, Germany\\ 
             \email{meus@tls-tautenburg.de} 
         \and
             University Leipzig, Faculty of Physics and Geosciences, 
             Linn\`estr. 5, D--04103 Leipzig, Germany
         \and
             Zentrum f\"ur Astronomie der Universit\"at Heidelberg,
             Astronomisches Rechen-Institut,
             M\"{o}nchhofstra\ss{}e 12-14, 69120 Heidelberg, Germany\\
             \email{hartmut@ari.uni-heidelberg.de}
             }

   \date{Received 10 August 2017; accepted ...}

 
  \abstract
   {} 
   {Using an accurate Gaia TGAS 25\,pc sample,
    nearly complete for GK stars,
    and selecting common proper motion (CPM) candidates
    from UCAC5, we search for new white dwarf (WD) companions
    around nearby stars with relatively small proper motions.}
   {For investigating known CPM systems in TGAS and for selecting
    CPM candidates in TGAS$+$UCAC5, we took into account the
    expected effect of orbital motion on the proper motion as well
    as the proper motion catalogue errors. Colour-magnitude
    diagrams (CMDs) $M_J/J-K_s$ and $M_G/G-J$ were used to verify
    CPM candidates from UCAC5. Assuming their common distance 
    with a given TGAS star, we searched for candidates that 
    occupied similar regions in the CMDs as the few known nearby 
    WDs (4 in TGAS) and WD companions (3 in TGAS$+$UCAC5). 
    CPM candidates with colours and absolute magnitudes corresponding
    neither to the main sequence nor to the WD sequence
    were considered as doubtful or subdwarf candidates.}
   {With a minimum proper motion of 60\,mas/yr, we 
    selected three WD companion candidates, two of which are
    also confirmed by their significant 
    parallaxes measured in URAT data,
    whereas the third may also be a chance alignment
    of a distant halo star with a nearby TGAS star.
    (angular separation of about 465\,arcsec).
    One additional nearby WD candidate was found from
    its URAT parallax and $GJK_s$ photometry.
    With \object{HD 166435 B} orbiting a well-known G1 star 
    at $\approx$ 24.6\,pc with a projected physical separation 
    of $\approx$700\,AU, we discovered one of the hottest 
    WDs, classified by us as DA2.0$\pm$0.2, 
    in the solar neighbourhood. We also
    found \object{TYC 3980-1081-1 B}, a strong cool WD companion 
    candidate around a recently identified new solar neighbour
    with a TGAS parallax corresponding to a distance of 
    $\approx$8.3\,pc and our photometric classification 
    as $\approx$M2 dwarf. This raises the question whether 
    previous assumptions on the completeness of the WD sample
    to a distance of 13\,pc were correct.}
   {}

   \keywords{Astrometry --
             Parallaxes ---
             Proper motions --
             binaries: general --
             white dwarfs --
             solar neighborhood
               }

\titlerunning{New nearby white dwarfs from TGAS+UCAC5/URAT}
   \maketitle
%

\section{Introduction}
\label{Sect_intro}

Most of the currently known nearby ($d$$<$25\,pc) stars were first suspected 
as solar neighbours because of their high proper motions. The lower limit of 
high proper motion catalogues has decreased with time, from 500\,mas/yr in the 
Luyten Half Second \citepads[LHS;][]{1998yCat.1087....0L} and 
180\,mas/yr in the New Luyten Two Tens \citepads[NLTT;][]{1995yCat.1098....0L} 
catalogues to 150\,mas/yr in the L{\'e}pine and Shara proper motion catalogue 
of Northern stars \citepads[LSPM-North;][]{2005AJ....129.1483L}, and now
to 40\,mas/yr in the publicly not yet available SUPERBLINK catalogue
\citepads{2017AAS...22915601L}. Only 15\% of about 3,300 stars within 25\,pc 
in the catalogue of nearby stars \citepads[CNS;][]{1995yCat.5070....0G}
have proper motions smaller than 180\,mas/yr, and for only 2\% they lie
below 40\,mas/yr. Meanwhile, the 25\,pc sample of the REsearch Consortium 
On Nearby Stars (RECONS\footnote{http://www.recons.org/}) has been 
improved with respect 
to the accuracy of the measured trigonometric parallaxes and contains about
4,000 objects in 3,000 systems \citepads{2015IAUGA..2253773H}. While the
25\,pc census of AFGK-type stars had already been completed by the 
Hipparcos \citepads{1997ESASP1200.....E} mission that surveyed all these
bright stars for their trigonometric parallaxes 
independent of their proper motions and colours, the fainter 
white dwarfs (WDs) and M-type dwarfs had to be preselected as targets for 
time-consuming ground-based parallax programmes by their colours 
and/or proper motions. According to the RECONS, there was an increase of 11\%
for WDs and 25\% for M dwarfs if one considers the immediate solar 
neighbourhood (the 10\,pc sample) in 2012 compared to its status in 2000.

A new unbiased survey that will help to finally complete the stellar 
(and partly also the substellar) census of the solar 
neighbourhood independent of proper motion limits is now being carried out 
by the Gaia mission \citepads{2016A&A...595A...1G}. A first relatively small, 
still incomplete and magnitude-limited ($V< 11.5$), subset of 
very accurate parallaxes was provided with the about two million stars in the
Tycho-Gaia Astrometric Solution \citepads[TGAS;][]{2016A&A...595A...4L}.
The TGAS catalogue is expected to include the majority 
of the GK-type 
stars within 25\,pc, whereas parallaxes of the brighter AF-type Solar 
neighbours and the fainter nearby WDs 
and the most frequent solar neighbours, the M-type stars, as well as most ML- 
and some of the T-type brown dwarfs \citepads[see][]{2017MNRAS.469..401S}
will only be provided in later Gaia data 
releases. The largest intermediate parallax survey between Hipparcos and Gaia
was provided by \citetads{2016AJ....151..160F}, who used data from the first
U.S. Naval Observatory Robotic Astrometric Telescope 
Catalog \citepads[URAT1;][]{2015AJ....150..101Z}. 
The full URAT Parallax Catalog \citepads[UPC;][]{2016arXiv160406739F}
contains more than 112,000 stars north of $\delta$$=$$-$13$^{\circ}$ 
available from \citetads{2016yCat.1333....0F} 
with significant parallax measurements,
including 53,500 stars, for which no previous parallaxes were available.
Note that the UPC is not biased towards high proper motion objects.

The positions of all stars published in the first Gaia data 
release \citepads[Gaia DR1;][]{2016A&A...595A...2G,2016A&A...595A...4L} 
were used to improve the
proper motion measurements of fainter stars in the 5th United States Naval 
Observatory CCD Astrograph Catalog\citepads[UCAC5;][]{2017AJ....153..166Z}.
The UCAC5 proper motions are of similar accuracy as those of TGAS 
(1-2\,mas/yr) for stars with $R$ magnitudes between 11 and 15. Out of all
107 million UCAC5 stars, 25 million have proper motion errors smaller 
than 2\,mas/yr. Therefore, we can search among the typically fainter UCAC5 
stars for objects that have a common proper motion (CPM) with a TGAS 
star. The CPM companions found in UCAC5 can then be assumed to lie at the 
same distances as their primary stars from TGAS, respectively. The CPM 
method has been usually applied to high proper motion 
stars, for which the errors were much smaller than the proper motion values
\citepads[e.g.][]{1997yCat.1130....0L,2014AJ....148...60L,2011ASPC..448.1375L}.
Very wide binaries with projected physical separations of $\gtrsim$1\,pc
were already analysed in TGAS data alone by \citetads{2017AJ....153..257O}
and
\citetads{2017MNRAS.472..675A}. 
Interestingly, apparent members
of open clusters, which lie at distances from the sun between 45\,pc and 
450\,pc, were found at very large separations (up to 15\,pc) from their 
cluster centres \citepads{2017A&A...601A..19G}, not only by their CPM status
but also according to their common parallaxes in TGAS.

Within the traditional CNS horizon of 25\,pc from the sun, 
\citetads{2017MNRAS.465.2849T} have found and analysed only four 
WDs directly observed in TGAS. In addition they identified 
nine such nearby TGAS stars, which have wide WD companions that were too 
faint to be included in TGAS. Compared to the total number of more than 180
currently known WDs with accurate parallaxes
within 25\,pc \citepads{2017AJ....154...32S}\footnote{According to
http://www.denseproject.com/25pc/ more than 20\% of these 
nearby WDs are components of binary/multiple systems.}, 
the TGAS numbers of directly and indirectly observed WDs are very small.
\citetads{2016MNRAS.462.2295H} list 232 WDs with trigonometric parallaxes or
spectroscopic distance estimates in their 25\,pc sample.
Despite of the obvious incompleteness of the TGAS 25\,pc sample with 
respect to low-mass stars and WDs, mostly fainter than the TGAS magnitude
limit, and although the UCAC5 does also not go deep enough to detect 
the coolest WDs 
at about 25\,pc,  we aimed at a CPM search for unknown nearby WD companions 
using the very accurate proper motions in both TGAS and UCAC5.
In particular, we wanted to extend the CPM method to those of the known
nearest stars that exhibit relatively small proper motions.

\section{Common proper motion of nearby wide binaries}
\label{Sect_cpm}

\subsection{The TGAS 25\,pc sample}
\label{Sect_tgas25}

We have selected 973 stars from TGAS whose parallaxes are larger than 40\,mas
and cross-identified this TGAS 25\,pc sample with the Two Micron 
All Sky Survey \citepads[2MASS;][]{2006AJ....131.1163S} after converting 
the TGAS coordinates with epoch 2015 to epoch 2000 taking into account 
the TGAS proper motions and using TopCat \citepads{2005ASPC..347...29T}.
A search radius of 4\,arcsec was used, as the 2MASS coordinates are given
for different epochs between 1997 and 2001. Applying the TGAS proper motions
to the 2MASS positions, we then converted the latter also to the epoch 2000.
The resulting angular distances between the epoch 2000 positions from both 
catalogues were small (median of 0.12\,arcsec, with only few values above
0.5\,arcsec and a maximum of 1.1\,arcsec).
All 973 stars were matched and provided with 2MASS $JHK_s$
magnitudes (except for one star lacking the $K_s$ magnitude). However,
only 871 out of the 973 stars have high-quality 2MASS photometry (quality 
flags ''AAA'').

The number of TGAS stars within 25\,pc corresponds to only a quarter of
the already mentioned RECONS 25\,pc sample. However, all TGAS stars within 
25\,pc have trigonometric parallax errors of less than 1\,mas 
(for $\approx$75\% the errors are even $<$0.5\,mas),
whereas the corresponding RECONS parallaxes were required to have errors
of less than 10\,mas. Hence the TGAS 25\,pc sample is highly incomplete
but very accurate
with respect to its astrometric measurements. 
The proper motion errors for TGAS stars within 25\,pc 
($\approx$88\% of them are Hipparcos stars)
are also much smaller than for other TGAS stars 
(for $\approx$80\% the errors are $<$0.2\,mas/yr).

\subsection{Effect of wide orbital motion on proper motions}
\label{Sect_pmo}

When we search for CPM objects to known nearby stars with accurately 
measured distances, 
we can estimate the expected influence of orbital motion on the proper 
motion difference between both components in a CPM pair. For simplicity
we assumed \textbf{(1)} a circular orbit in the plane of the sky, 
\textbf{(2)} an orbital radius corresponding to the projected 
physical separation (in AU) that can be computed as the product 
of the angular separation between both components (in arcsec) 
and the assumed common distance (in pc), 
and \textbf{(3)} a total mass of the system of 1.5 solar masses. 
With these assumptions, and following Kepler's third law, we can
compute the orbital period and velocity and transform the latter to a 
proper motion, again using the known distance. This proper motion effect
due to 
orbital motion, $pmo$, is the expected maximum proper motion 
difference between the components of a nearby wide binary. This
effect increases with smaller distances and separations. For a wide
binary located at 25\,pc from the sun, separations of 3600\,arcsec, 
60\,arcsec, and 6\,arcsec lead to $pmo$ of about 1\,mas/yr, 8\,mas/yr, 
and 25\,mas/yr, respectively. With equal separations, the corresponding 
$pmo$ are already about 4 times larger at 10\,pc, and about 11 times
larger at 5\,pc, respectively.

\begin{figure}
   \centering
   \includegraphics[width=\hsize]{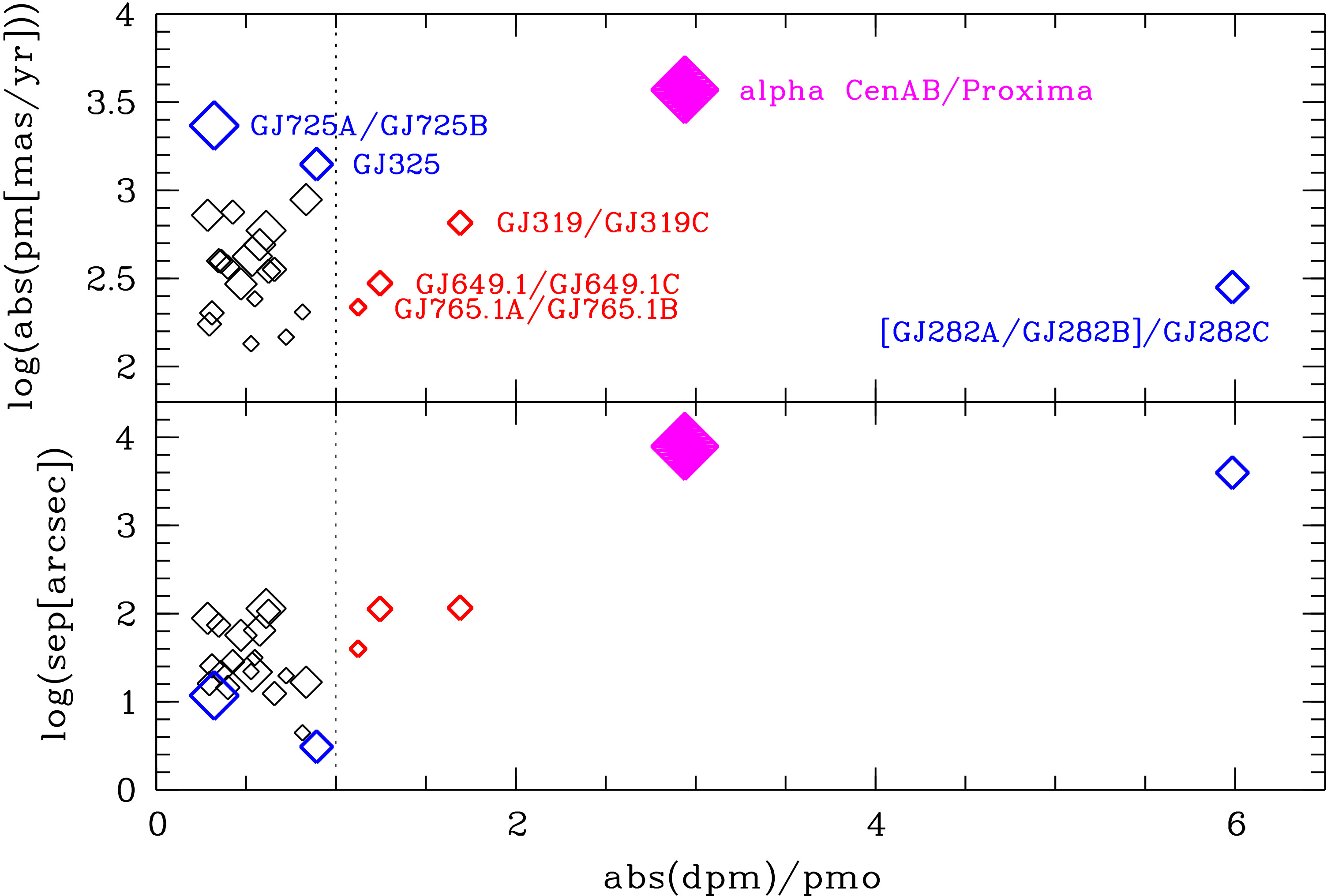}
      \caption{Wide binaries and multiple systems 
               with common parallaxes and relatively
               small angular separations (see text)
               in the TGAS 25\,pc sample. Shown are their total proper
               motions \textbf{(top)} and angular separations 
               \textbf{(bottom)} as a function of the ratio 
               of the absolute value of their
               total proper motion difference $dpm$ (see
               Eq.~\ref{Eq_2}) 
               to the estimated effect caused by orbital motion ($pmo$). 
               The dotted line indicates the expected maximal ratio of 1. 
               Systems discussed in the text 
               are labeled in the top panel and marked by 
               thick red and blue open symbols.
               For comparison we added the values
               for \object{Proxima} with respect to \object{alpha CenAB} 
               (filled magenta symbols), 
               which are not included in TGAS. 
               The symbol sizes increase with parallax. 
              }
         \label{Fig_ratio_dpm_pmo}
   \end{figure}

Before we used $pmo$ as a criterion for our CPM search in TGAS$+$UCAC5 
data (see Sect.~\ref{Sect_tgasucac5wdcpm}), we investigated the known CPM
pairs in the TGAS 25\,pc sample alone. Here we have the advantage that
we can find physical pairs of stars from their common distance and small 
angular separation and then check the agreement in their proper motions.
As the TGAS proper motion errors are typically very small
($\ll$2\,mas/yr) and can be neglected, 
we used in this case the total proper motion differences 
(Eq.~\ref{Eq_2}), leading also to a clearer presentation in 
Fig.~\ref{Fig_ratio_dpm_pmo}. However, in our TGAS$+$UCAC5 CPM search 
(Sect.~\ref{Sect_tgasucac5wdcpm}), we applied the $pmo$ criterion to the 
individual proper motion components (in RA and DE) taking into account
their typically larger (and sometimes rather different) UCAC5 errors.

We searched for wide 
binaries and multiple systems 
within the TGAS 25\,pc sample 
with angular separations of up to several degrees. First, we allowed only 
for parallaxes that agree within their TGAS errors and 
found 18 CPM pairs 
with total proper motion differences smaller than $pmo$. 
Their angular separations ranged between 
4-120\,arcsec, 
their parallaxes between 45-114\,mas, and 
their total proper motions between 130-900\,mas/yr. 
Allowing for two times larger proper motion differences, 
we found three more 
CPM pairs falling in the same ranges of angular separations, parallaxes,
and total proper motions, respectively. These three 
CPM pairs correspond to well-known systems of G-type stars
(\object{GJ 765.1A}/\object{GJ 765.1B}),
K dwarfs 
(\object{GJ649.1}(=\object{CCDM J16548+4722AB)}/\object{GJ649.1C}),
and of early-M dwarfs
(\object{GJ 319}(=\object{CCDM J08427+0933AB})/\object{GJ 319C}),
respectively (thick red 
open 
symbols in the central parts of the two
panels of Fig.~\ref{Fig_ratio_dpm_pmo}). 
Finally, as shown in Fig.~\ref{Fig_ratio_dpm_pmo}, 
we arrived at 24 CPM systems, 
when we conservatively allowed for 10\% differences in the TGAS parallaxes.
Interestingly, all the latter three additional cases fall outside of at least
one of the above given ranges and occupy edge regions in
the two panels of Fig~\ref{Fig_ratio_dpm_pmo} 
(objects marked by thick blue 
open
symbols). One of these pairs (\object{GJ 325} = \object{CCDM J08555+7048AB}) 
consists of two early-M dwarfs with
a very small angular separation of 3\,arcsec and 
a large total proper motion of about 1400\,mas/yr.
The second
pair (\object{GJ 725A}/\object{GJ 725B}), 
of slightly cooler mid-M dwarfs,
has also a relatively small angular separation 
of 11\,arcsec, a large parallax of $>$280\,mas, and 
very large total proper motion of about 2300\,mas/yr.
These two pairs fall in the lower and upper left parts of the two panels in 
Fig.~\ref{Fig_ratio_dpm_pmo}, respectively.

The third system, consisting of three TGAS sources, the wide binary 
\object{GJ 282A}/\object{GJ 282B} and its much wider companion
\object{GJ 282C} (shown by one symbol at the right edge of the top and 
bottom panels of Fig.~ref~\ref{Fig_ratio_dpm_pmo}, respectively),
does not stand out by its parallax or proper motion
but has an extremely large angular separation of about 3900\,arcsec. This
is much larger than 
the angular separations of 
all other wide binaries in the TGAS 25\,pc sample. 
At a distance of about 14\,pc it leads to a projected physical separation of
about 56000\,AU, even larger than the about 10000\,AU between
\object{alpha CenAB} and \object{Proxima}. 
The early-M dwarf \object{GJ 282C} was studied 
with respect to the 
K-type binary 
\object{GJ 282A}/\object{GJ 282B} and found to 
rank among the widest physical companions by \citetads{2009ApJ...706..343P}.
As our nearest neighbours, \object{Proxima} and \object{alpha CenAB},
represent another prominent case of an extremely wide but bound system 
\citepads{2017A&A...598L...7K}, we added the corresponding values in
Fig.~\ref{Fig_ratio_dpm_pmo}. 
These 
objects are not in TGAS, so we used
their data as listed in \citetads{2017A&A...598L...7K}. 
For [\object{GJ 282A}/\object{GJ 282B}]/\object{GJ 282C} 
we find ratios between the 
total proper motion differences 
and our estimated proper motion effect $pmo$ of about 6.
These ratios exceed even those of \object{alpha CenAB}/\object{Proxima}. 
Such very large ratios may indicate that our simple 
assumptions for computing $pmo$ are not correct and/or that these systems
are in the process of 
dynamical disintegration \citepads{2009ApJ...706..343P}.

\subsection{The role of chance alignments}
\label{Sect_chance}

The larger the separation between the components of a wide binary
candidate and the smaller their assumed CPM, the more likely is a chance
alignment of unrelated objects. 
The proper motion errors do also play an important role in
the correct identification of CPM pairs, in particular if they are of the
same order as the expected proper motion differences due to orbital motion
(Sect.~\ref{Sect_pmo}) and/or if they are not much smaller than the
proper motion values. \citetads{2007AJ....133..889L} investigated 
CPM companions of Hipparcos primaries in the LSPM-North catalogue
and excluded chance alignments using an empirial relation between 
the separation ($sep$), total proper motion difference ($dpm$), and 
the total proper motion ($pm$):
\begin{equation}
\label{Eq_1}
\frac{sep*dpm}{(pm/pm_{min})^{3.8}}<1,
\end{equation}
where $sep$ was given in arcsec, whereas $dpm$ and $pm$ were here in 
arcsec/yr (but are in 
mas/yr 
throughout this paper),
and $pm_{min}$ was set to 0.15\,arcsec/yr, the lower proper motion limit
of the LSPM-North catalogue.
The total proper motion difference was computed as:
\begin{equation}
\label{Eq_2}
dpm=\sqrt{(dpmRA^2+dpmDE^2)}.
\end{equation}
In our search for new WD companions 
(Sect.~\ref{Sect_tgasucac5wdcpm}), 
we applied several criteria, starting with a lower proper motion limit,
selecting a maximum angular separation (similar to the limit used by
\citetads{2007AJ....133..889L}), taking into account the 
proper motion errors, and the ratio of
the proper motion differences to the $pmo$ of each CPM candidate. In our
final selection we considered only ratios below 1.5 so that we may have
excluded some possible extremely wide binaries, similar to the cases
described in Sect.~\ref{Sect_pmo}. For an evaluation of our new WD CPM 
candidates and known WD CPM companions we checked, whether they fulfil
the condition of Eq.~\ref{Eq_1}.

\section{Nearby TGAS/UCAC5 WD companion candidates}
\label{Sect_tgasucac5wdcpm}

\subsection{CPM search criteria}
\label{Sect_cpmcriteria}

Our cross-matching of the TGAS 25\,pc sample with the UCAC5 with a large
search radius of 3600\,arcsec yielded more than 7 million UCAC5 stars, 
on the average about 7200 field
stars per TGAS star. For each field star, we determined
the $pmo$ effect that we expected if it were a wide binary companion of 
the TGAS star (see Sect.~\ref{Sect_pmo}). When we consider the proper motion
differences $dpm$ (TGAS-UCAC5) of potential CPM pairs, we must also take 
into account the proper motion errors $epm$ in the TGAS 25\,pc 
sample and in UCAC5. Even after excluding UCAC5 stars with large proper 
motion errors, the TGAS proper motions were still about ten times more 
accurate than those of the remaining UCAC5 stars. This is due to the fact
that the TGAS 25\,pc sample consists mostly of Hipparcos stars, for which
the TGAS was much more accurate than for Tycho stars. Only the 120
non-Hipparcos stars in the TGAS 25\,pc sample show a similar peak in 
the proper motion error distribution as in UCAC5. However, the tail
of their error distribution contains only about 10 stars with 
errors $>$3\,mas/yr with a maximum at about 6\,mas/yr. Therefore, we
did not exclude potential CPM primaries based on TGAS proper motion errors. 
We selected candidate TGAS/UCAC5 CPM pairs with the following criteria:

\begin{itemize}
\item TGAS and UCAC5 total proper motion $>$60\,mas/yr,
\item TGAS-UCAC5 separation $<$1800\,arcsec,
\item $epm_{UCAC5}<$6\,mas/yr (in RA and DE components),
\item $|dpm|<1.5*(pmo+epm_{TGAS}+epm_{UCAC5})$ (in RA and DE).
\end{itemize}

When we tried to change the first three limits towards smaller total proper
motions, larger separations, and larger UCAC5 proper motion errors,
respectively, the number of CPM candidates increased. However, as their 
total proper motions were generally smaller, 
it happened that several CPM candidates were found 
for a given TGAS primary, in particular for very nearby TGAS stars. 
The additional candidates also showed unrealistic
absolute magnitudes (assuming a common distance with the primary) with
their given colours in near-infrared and optical to near-infrared 
colour-magnitude diagrams (CMDs). 
The factor of 1.5 in the fourth selection criterion
was a compromise justified by the fact that several known wide binaries 
among the nearby CPM pairs in TGAS alone and in TGAS/UCAC5 would have been 
excluded with a factor of 1.0, whereas factors of 2.0-2.5 
led again to more doubtful candidates. 

After applying the first selection criterion, the number of 
CPM candidates reduced to about 86,000 ($\approx$1\% of the original
number of selected field stars). With the
following two criteria, it further reduced to about 22,000 and 11,000, 
respectively. The fourth and decisive criterion led to 
only 72 ($\approx$0.001\%) CPM candidates, most of which turned out to 
be main sequence (MS) stars
(see Figs.~\ref{Fig_nircmd} and \ref{Fig_optnircmd}).

\subsection{WD companion candidates in CMDs}
\label{Sect_wdcmd}

\begin{figure}[t!]
   \centering
   \includegraphics[width=\hsize]{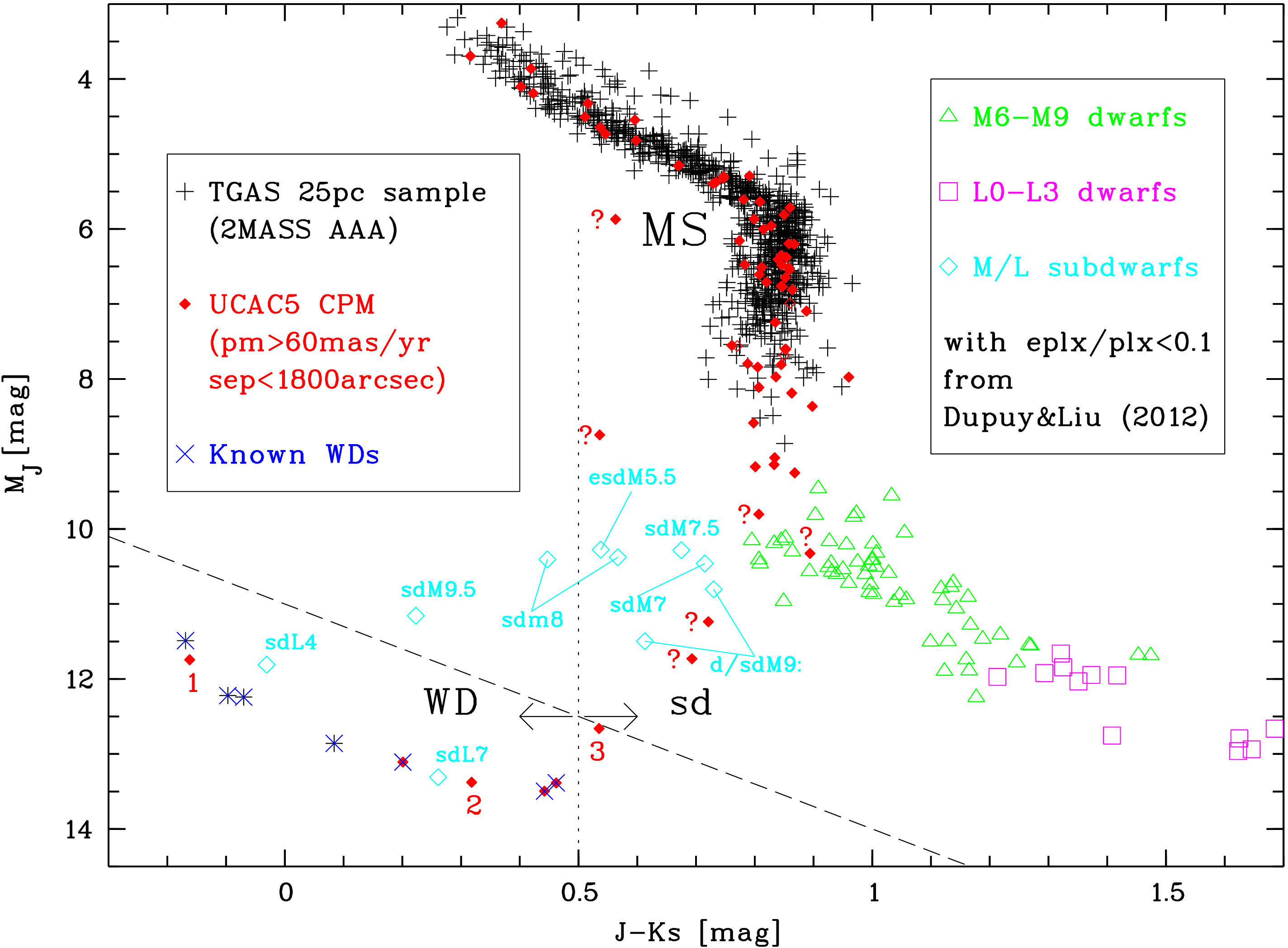}
      \caption{Near-infrared 2MASS CMD for stars in
               the TGAS 25\,pc sample with photometric quality 'AAA'
               in the 2MASS catalogue (black plus signs)
               and for selected CPM companions (see text)
               from UCAC5 (filled red lozenges). Their absolute magnitudes
               $M_J$ are based on TGAS parallaxes. Four known WDs in
               TGAS and three known WD companions in UCAC5 are marked
               by blue crosses. Our WD candidates
               (see Eqs.~\ref{Eq_3} and \ref{Eq_4})
               fall below both dashed lines in
               this figure and in Fig.~\ref{Fig_optnircmd}, respectively.
               The dotted line shows the dividing line
               at $J-K_s$$=$$+$0.5 between WDs and cool subdwarfs (sd)
               found by
               \citetads{2017AJ....154...32S}. Also shown are ''normal''
               (not flagged as atypical, and not known as close binaries)
               late-M (green open triangles) and
               early-L dwarfs (magenta open squares) as well
               as late-M and L subdwarfs (cyan open lozenges;
               marked by their spectral types) from
               \citetads{2012ApJS..201...19D} based on their parallaxes
               with relative errors less than 10\%. 
               Three new WD companion
               candidates are numbered. All CPM candidates
               with angular separations $>$900\,arcsec
               are labelled with question marks.
              }
         \label{Fig_nircmd}
   \end{figure}

After selecting among the 973 stars of the TGAS 25\,pc sample only
those 871 with accurate 2MASS photometry, 
the near-infrared CMD in Fig.~\ref{Fig_nircmd}
(black plus signs) shows the MS 
populated 
from (a few late-F and) early-G to early-M spectral types
and a small WD sequence of four isolated WDs,
\object{WD 1142-645} (DQ6.4),
\object{WD 1647+591} (DAV4.1), 
\object{WD 1917-077} (DBQZ4.9), 
and \object{WD 2117+539} (DA3.6),
as already mentioned by \citetads{2017MNRAS.465.2849T}
\citepads[spectral types are taken from][]{2016MNRAS.462.2295H}.
Figure~\ref{Fig_nircmd} also shows the CPM
companions identified in our search (filled red lozenges), 
where we assume that the parallax
of a CPM companion is the same as that of its TGAS primary.
With our CPM companion search aiming at WD companions we did not expect 
to find late-M dwarfs (and brown dwarfs with late-M, L or T spectral types) 
with $M_J$$>$10, as even the
nearest of these objects are too faint to be included in UCAC5.
Consequently, the UCAC5 CPM companions only slightly extend the MS
towards fainter magnitudes. For comparison we show in Fig.~\ref{Fig_nircmd}
that late-M and early-L dwarfs occupy the lower right part of the 
near-infrared CMD as derived from accurate parallaxes 
by \citetads{2012ApJS..201...19D}.

From nine known WD companions of TGAS stars within 25\,pc
\citepads{2017MNRAS.465.2849T}, only three are included in UCAC5
(\object{WD 0433+270}, \object{WD 0751-252}, and \object{WD 2154-512}).
All three were found by our CPM selection criteria. They extend the
small WD sequence in Fig.~\ref{Fig_nircmd} (red filled lozenges
overplotted by blue crosses) to the red. Their angular
separations range between 28\,arcsec and 400\,arcsec.
The other six WD companions that are not in UCAC5 have
smaller angular separations (7-17\,arcsec) from their bright primaries,
which 
may have prevented 
their detection on the UCAC images,
and two of them may also have been too faint ($V$$>$16) to be observed
within the UCAC survey.

The data of known and candidate nearby ($d<25$\,pc) TGAS/UCAC5 CPM
pairs including a WD companion, all found with the selection criteria
described in Sect.~\ref{Sect_cpmcriteria}, are listed in
the upper part of
Table~\ref{Tab_cpm}.
Two of the three new WD companion candidates have relatively small
angular separations, whereas the third, \object{HD 35650 B}, looks with
its large angular separation at approximately the same distance
similar to the known
very wide CPM companion \object{SCR J0753-2524}. Consequently, in these
two pairs the $pmo$ effect is relatively small compared to the proper
motion errors.

   \begin{figure}[t!]
   \centering
   \includegraphics[width=\hsize]{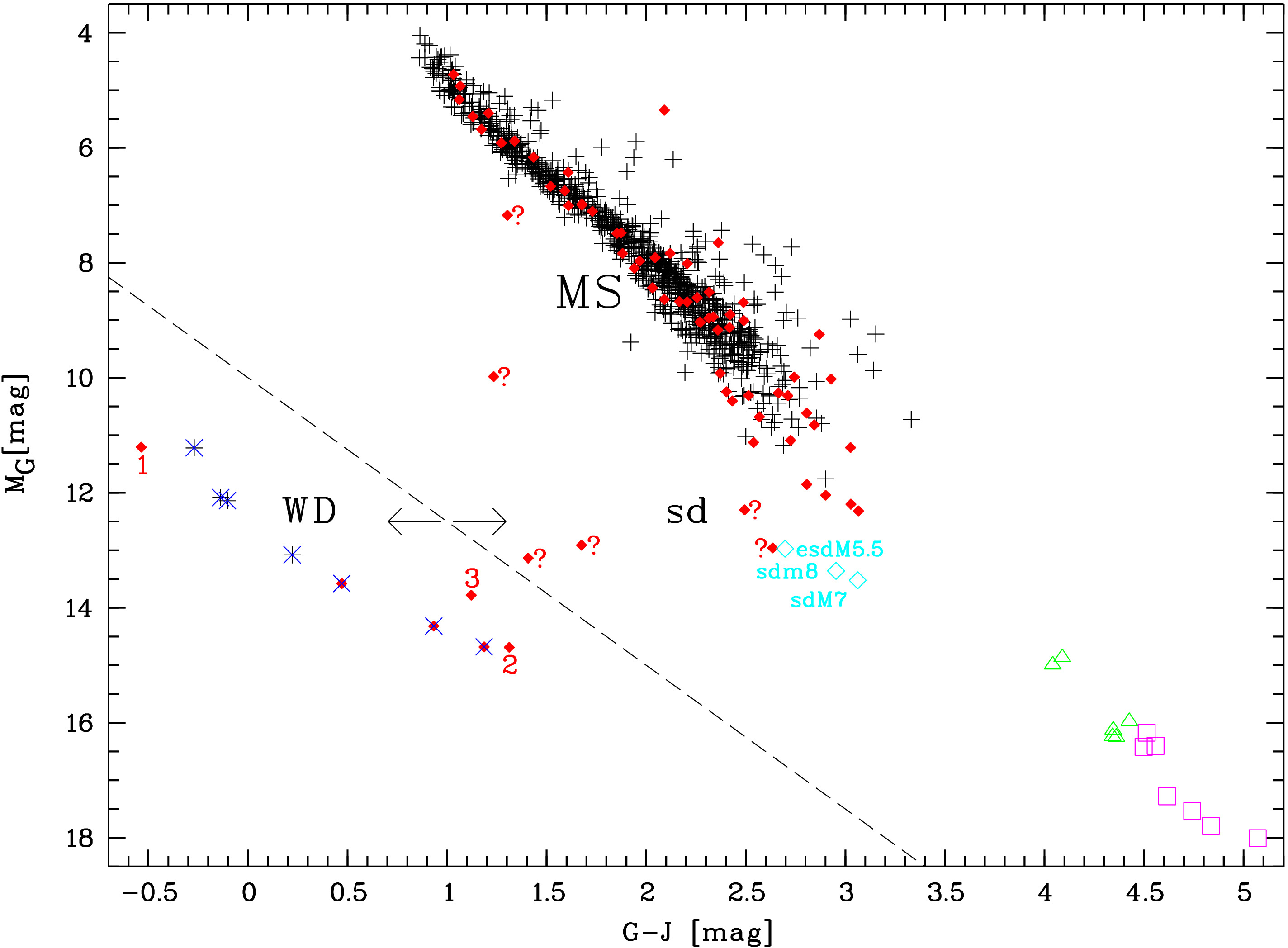}
      \caption{Optical (Gaia) to near-infrared (2MASS) CMD 
               with the same stars (and symbols) as in Fig.~\ref{Fig_nircmd}.
               Only part of the late-M and early-L dwarfs, and of the
               M/L subdwarfs from \citetads{2012ApJS..201...19D} could
               be identified in Gaia (see text). As in Fig.~\ref{Fig_nircmd},
               the absolute Gaia magnitudes $M_G$ of TGAS stars and UCAC5 CPM 
               candidates are based on TGAS parallaxes, whereas those of
               the comparison objects are based on the parallaxes 
               from \citetads{2012ApJS..201...19D}. 
               Objects located below the 
               dashed lines in this figure and in Fig.~\ref{Fig_nircmd}
               were considered as WDs.
              }
         \label{Fig_optnircmd}
   \end{figure}

From Fig.~\ref{Fig_nircmd} it is obvious that the WD companion candidates '1' 
and '2' fit well into the sequence of known WDs, whereas candidate '3' 
lies a bit off and just right of the WD/subdwarf
dividing line described by \citetads{2017AJ....154...32S}. Doubtful 
candidates marked with '?', which had the largest angular separations 
($\approx$1000-1750\,arcsec)
but relatively small 
(60-90\,mas/yr) 
total proper motions, are in this
near-infrared CMD either much redder or brighter than the WD sequence.
However, because of their small proper motions and the spectral
classification (in SIMBAD) of their apparent primaries as normal MS stars,
we can not consider these doubtful CPM companions as nearby subdwarf 
candidates. With their large separations, it is much more likely that 
these are chance alignments of distant MS stars (with distances 
in the range of $\approx$50-500\,pc according to their $G-J$ colour and
apparent $G$ magnitudes) with the nearby TGAS stars. In fact, if we 
compute the ratio according to Eq.~\ref{Eq_1}, 
replacing the $pm_{min}$ used by \citetads{2007AJ....133..889L} with
our lower proper motion limit $pm_{min}$=0.06\,arcsec/yr, we get 
for these six doubtful CPM pairs values 
between 1.4 and 5.3, indicating chance alignments. The tangential 
velocities derived from the proper motions and the estimated larger 
photometric distances of these rejected CPM companions
are mostly in the range of 20-80\,km/s, 
typical of the Galactic thin disk population. Only for one object,
which can be identified as \object{2MASS 06233146-5952448},
located just above the dashed line in Fig.~\ref{Fig_optnircmd},
the resulting tangential velocity was larger ($\approx$150\,km/s)
so that 
this might be 
a distant (K-type) subdwarf 
(thick disk object), unrelated to our Solar neighbourhood TGAS sample.

For candidate '3', this ratio is also large (2.1) indicating
a probable chance alignment. However, its separation is only moderately
large (about 465\,arcsec) and it lies much closer to the WD sequence.
Therefore, we consider this object as a weak WD companion candidate.
Interestingly, its relatively blue colour ($G-J\approx+1.1$) and faint 
magnitude ($G\approx$15) lead to an alternative photometric distance of
$\approx$800\,pc and tangential velocity of $\approx$260\,km/s, if
it is unrelated. In this case, this would be a distant G-type subdwarf 
candidate, possibly belonging to the Galactic halo population.
For all other CPM candidates, including known WD companions and
the WD companion candidates '1' and '2', 
the ratio from Eq.~\ref{Eq_1} was always smaller than 0.4,
with a mean value of 0.02, confirming their real CPM status.

Concerning the location of candidate '3' close to the WD/subdwarf 
dividing line (dotted line in Fig.~\ref{Fig_nircmd}) we note that 
the $J-K_s$ colours used in \citetads{2017AJ....154...32S} were not directly 
measured in but transformed to the 2MASS system. The position 
of this dividing line is questionable in view of
the wide-spread $J-K_s$ colour distribution of the shown M/L subdwarfs 
and obviously not valid for the latest-type (coolest) known subdwarfs,
\object{SSSPM J1013-1356} (sdM9.5), \object{2MASS J16262034+3925190} (sdL4), 
and \object{2MASS J05325346+8246465} (sdL7).
The remaining seven M/L subdwarfs seem 
to represent better
the 60\,pc cool 
subdwarf sample (presumably dominated by sdK and early-sdM subdwarfs) that 
was used by \citetads{2017AJ....154...32S} in finding the $J-K_s=+0.5$
dividing line. 
We considered objects as WD candidates if they were falling 
below two colour-magnitude limits (shown by the dashed lines 
in Figs.~\ref{Fig_nircmd}  and \ref{Fig_optnircmd}):
\begin{equation}
\label{Eq_3}
M_G>10.0+(2.5*(G-J)),
\end{equation}
\begin{equation}
\label{Eq_4}
M_J>11.0+(3.0*(J-K_s)).
\end{equation}

%
%
\begin{table*}
\caption{Astrometry and photometry of known and candidate ('1-3') WD companions
and their primaries in nearby TGAS/UCAC5 CPM pairs and of additional known WDs with no previous
trigonometric parallaxes (in SIMBAD) and one WD candidate ('4') found in the UPC}.
\label{Tab_cpm}
\centering
\fontsize{7pt}{0.90\baselineskip}\selectfont
\begin{tabular}{@{}ll@{}rclrc@{}crcl@{}c@{}r@{}}     
\hline\hline
Name & RA\tablefootmark{a} & DE\tablefootmark{a} & $plx_{TGAS}$          & $plx_{UPC}$ & $pm$RA & $pm$DE & $pmo$ & $G$\tablefootmark{a}   & $J$\tablefootmark{b} & $J-K_s$\tablefootmark{b} &SpT & Ref \\
(WD name/cand)     &          &             & (Sep, PA)\tablefootmark{a}             &  & & & ($|dpm|$)\tablefootmark{f} \\
     & [degrees]   & [degrees]   & [mas]                 & [mas] & [mas/yr]                   & [mas/yr] & [mas/yr]     & [mag] & [mag] & [mag]      & \\
     &             &             & ([arcsec,$^{\circ}$]) & & & & ([mas/yr]) \\
\hline
\object{GJ 171.2 A} & 069.202096 & $+$27.131577 & 57.2$\pm$0.3 & 65.1$\pm$9.0  & $+$232.1$\pm$0.1\tablefootmark{c} & $-$147.7$\pm$0.1\tablefootmark{c} & 9     &  7.64 &  5.95 & $+$0.71 & K3IVke & 3 \\
\object{GJ 171.2 B} & 069.188131 & $+$27.163694 & (124.0,339)  & 80.1$\pm$11.0 & $+$230.8$\pm$3.0\tablefootmark{d} & $-$146.6$\pm$2.9\tablefootmark{d} & (1,1) & 15.53 & 14.60 & $+$0.46 & DA9.0  & 1 \\
(= \object{WD 0433+270}) & & & & & $+$229.7$\pm$6.1\tablefootmark{e} & $-$142.9$\pm$6.0\tablefootmark{e} \\
\object{Ross 429}       & 118.543958 & $-$25.302335 & 56.2$\pm$0.3 & - & $-$300.8$\pm$0.1\tablefootmark{c} & $+$201.0$\pm$0.1\tablefootmark{c} & 5      &  8.97 &  7.02 & $+$0.85 & M0    & 4 \\
\object{SCR J0753-2524}\tablefootmark{i} & 118.484536 & $-$25.399551 & (399.8,209)  & - & $-$286.4$\pm$5.2\tablefootmark{d} & $+$205.5$\pm$5.3\tablefootmark{d} & (14,4) & 15.93 & 14.75 & $+$0.44 & DA9.8 & 1 \\
(= \object{WD 0751-252}) \\
\object{GJ 841 A} & 329.421448 & $-$51.007760 & 66.1$\pm$0.5 & - & $-$35.3$\pm$0.2\tablefootmark{c} & $-$379.8$\pm$0.2\tablefootmark{c} & 25      &  9.24 &  6.75 & $+$0.87 & M2Ve  & 5 \\
\object{GJ 841 B} & 329.409514 & $-$51.010412 & (28.7, 251)  & - & $-$46.5$\pm$1.2\tablefootmark{d} & $-$395.5$\pm$1.1\tablefootmark{d} & (11,16) & 14.48 & 14.01 & $+$0.20 & DQ8.3 & 1 \\
(= \object{WD 2154-512}) \\
\object{HD 166435} & 272.339436 & $+$29.951968 & 40.7$\pm$0.2 & 46.1$\pm$8.8 & $+$71.9$\pm$0.1\tablefootmark{c} & $+$61.1$\pm$0.1\tablefootmark{c} & 12     &  6.62 &  5.69 & $+$0.37 & G1IV  & 3 \\
\object{HD 166435 B}& 272.331393 & $+$29.956100 & (29.2,301)   & 50.4$\pm$3.3 & $+$64.8$\pm$1.2\tablefootmark{d} & $+$72.4$\pm$1.2\tablefootmark{d} & (7,11) & 13.16 & 13.70 & $-$0.16 & DA2.2$\pm$0.2 & 2 \\
(= '1') & & & & & $+$67.4$\pm$2.4\tablefootmark{e} & $+$71.5$\pm$2.4\tablefootmark{e} \\
\object{TYC 3980-1081-1}   & 327.909017 & $+$59.294450 & 120.6$\pm$1.0 & 154.7$\pm$12.1 & $-$70.8$\pm$2.5\tablefootmark{c} & $+$76.1$\pm$2.4\tablefootmark{c} & 84       &  9.20 &  6.53 & $+$0.88 & $\approx$M2 & 2 \\
\object{TYC 3980-1081-1 B} & 327.916408 & $+$59.292935 & (14.6,112)    & 131.0$\pm$4.4  & $-$87.2$\pm$2.2\tablefootmark{d} & $-$31.2$\pm$2.0\tablefootmark{d} & (16,107) & 14.28 & 12.97\tablefootmark{g} & $+$0.32\tablefootmark{g} & $\approx$DA10? & 2 \\
(= '2') & & & & & $-$79.5$\pm$3.3\tablefootmark{e} & $-$17.3$\pm$3.2\tablefootmark{e} \\
\object{HD 35650}   & 081.125938 & $-$38.969890 & 57.4$\pm$0.3 & - & $+$43.1$\pm$0.1\tablefootmark{c} & $-$57.3$\pm$0.1\tablefootmark{c} & 5     &  8.43 &  6.70 & $+$0.78 & K6Vke & 6 \\
\object{HD 35650 B} & 081.252385 & $-$39.053688 & (464.9,131)  & - & $+$46.3$\pm$1.9\tablefootmark{d} & $-$48.8$\pm$2.0\tablefootmark{d} & (3,9) & 14.99 & 13.87 & $+$0.53 & $\approx$DA9? & 2 \\
(= '3') \\
\hline
\object{GJ 3285}        & 066.473476 & $+$12.195784 & - & 58.0$\pm$8.0 &  $-$94.6$\pm$6.4\tablefootmark{e} &  $-$227.9$\pm$6.4\tablefootmark{e} & - & 15.21 & 14.49 & $+$0.24 & DC8.2 & 1 \\
(= \object{WD 0423+120} &  \\
\object{GJ 275.2 B}\tablefootmark{h}     & 112.695736 & $+$48.168756 & - & 89.6$\pm$4.2 & $-$216.6$\pm$4.0\tablefootmark{e} & $-$1274.9$\pm$3.8\tablefootmark{e} & - & 14.84 & 13.08 & $+$0.33 & DA10.0+DA10.1 & 1 \\
(= \object{WD 0727+482} &  \\
\object{GJ 1098}        & 113.379072 & $+$64.156548 & - & 54.3$\pm$8.2 &  $+$50.9$\pm$6.8\tablefootmark{e} &  $-$260.5$\pm$5.9\tablefootmark{e} & - & 15.91 & 14.81 & $+$0.43 & DA11.1 & 1 \\
(= \object{WD 0728+642} & \\
\object{GJ 4165}\tablefootmark{j}        & 312.278706\tablefootmark{e} & $+$37.471154\tablefootmark{e} & - & 63.1$\pm$2.9 & $+$167.1$\pm$2.9\tablefootmark{e} & $+$154.3$\pm$2.9\tablefootmark{e} & - & - & 13.30 & $-$0.13 & DA3.6 & 1 \\
(= \object{WD 2047+372} \\
\object{UPC 72924}      & 277.995093 & $+$46.974870 & - & 40.2$\pm$6.6 &  $+$61.5$\pm$4.6\tablefootmark{e} &   $+$38.4$\pm$4.4\tablefootmark{e} & - & 15.05 & 14.52 & $+$0.14 & $\approx$DA7? & 2 \\
(= '4') \\
\hline
\end{tabular}
\tablefoot{\fontsize{7pt}{0.90\baselineskip}\selectfont
Gaia coordinates are for (J2000, epoch 2015.0) and were
rounded to 0.000001 degrees. Although the TGAS data may be more accurate, all
parallaxes and their errors were rounded to 
0.1\,mas, proper motions and their errors to 0.1\,mas/yr. The absolute 
values of measured proper motion differences $dpm$ and the estimated effect 
because of orbital motion $pmo$ were rounded to 1\,mas/yr. Magnitudes and 
colours were rounded to 0.01\,mag. Spectral types with question marks 
are only guesses based on $G-J$ and $M_G$ (Figs.~\ref{Fig_optnircmd} and \ref{Fig_optnirupc}). 
Further notes on the data:
\tablefoottext{a}{Gaia DR1,}
\tablefoottext{b}{2MASS,}
\tablefoottext{c}{TGAS,}
\tablefoottext{d}{UCAC5,}
\tablefoottext{e}{UPC,}
\tablefoottext{f}{two values are given for RA and DE, respectively,}
\tablefoottext{g}{$J$ magnitude with very poor goodness-of-fit quality of the profile-fit photometry,}
\tablefoottext{h}{An accurate parallax of 87.4$\pm$0.5\,mas measured for this close binary by \citetads{2015ASPC..493..501N},}
\tablefoottext{i}{\citetads{2009AJ....137.4547S} measured a parallax of 56.54$\pm$0.95\,mas,}
\tablefoottext{j}{This object is not plotted in Figs.~\ref{Fig_nirupc} and \ref{Fig_optnirupc}, since it was not identified in Gaia DR1.}
}
\tablebib{\fontsize{7pt}{0.90\baselineskip}\selectfont
(1) \citetads{2016MNRAS.462.2295H};
(2) this paper;
(3) \citetads{2003AJ....126.2048G};
(4) \citetads{2014MNRAS.443.2561G};
(5) \citetads{2006A&A...460..695T};
(6) \citetads{2006AJ....132..161G}.
}
\end{table*}

Contrary to 
the WD/subdwarf confusion in a near-infrared $M_J/J-K_s$ diagram
(Fig.~\ref{Fig_nircmd}) we expect a clearer separation of the MS, subdwarf 
and WD sequences in an optical to near-infrared CMD. As Gaia provides accurate
magnitudes, albeit in a very broad optical band, we studied the TGAS 25\,pc 
sample and our UCAC5 CPM candidates together with the comparison objects 
from \citetads{2012ApJS..201...19D} in an $M_G/G-J$ diagram 
(Fig.~\ref{Fig_optnircmd}). Unfortunately, only three out of ten M/L subdwarfs
could be identified in Gaia DR1 (with Gaia magnitudes $16.3<G<17.7$).
This low identification rate may be due to the very large proper
motions, probably problematic for Gaia DR1, of these ultracool subdwarfs. 
Among the comparison late-M and early-L dwarfs, we identified 6 out of 58 
(only 10\%; $15.8<G<17.7$) and 7 out of 12 (58\%; $18.5<G<20.6$) objects 
in Gaia DR1, respectively. 
The very low identification rate of the relatively bright and nearby
late-M dwarfs may be caused by their large 
parallaxes 
and proper motions 
and the selection criteria for problematic sources in Gaia DR1.
The much fainter early-L dwarfs have typically smaller parallaxes and 
proper motions so that their identification rate is similar to that 
found for all L and T dwarfs in 
Gaia DR1 \citepads[45\%;][]{2017MNRAS.469..401S}. 

The MS in Fig.~\ref{Fig_optnircmd} shows a large gap at the expected 
position of early- to mid-M dwarfs. 
Three labelled late-M subdwarfs form a
parallel sequence shifted from the MS by $\approx$0.5\,mag to the
blue. Some of the doubtful CPM candidates marked with '?' seem to
fall in the same region, but we consider them as chance alignments
rather than nearby subdwarf candidates (see above).
The WD sequence is also 
almost parallel 
with respect to the MS, 
but shows a large blue-shift of $\approx$3.0\,mag.
The new WD candidates '1' and '2' are now located at the blue and red end 
of the WD sequence,
suggesting
a hot and a cool WD, respectively.
The two L-type subdwarfs, which were falling right in the WD sequence in
the near-infrared CMD (Fig.~\ref{Fig_nircmd}) have no Gaia magnitudes yet,
but we expect them to fall in the lower right corner 
of Fig.~\ref{Fig_optnircmd}, well separated from the WDs in $G-J$
colour. 

Note that all the M/L subdwarfs from \citetads{2012ApJS..201...19D}
are fainter than the UCAC5 magnitude limit of $R\approx16$\,mag.
The same is true for the MS late-M and early-L dwarfs.
As already discussed above, the Gaia magnitude intervals for the
M/L subdwarfs and the MS late-M/early-L dwarfs are 
almost identical,
and their UCAC5 magnitudes fall in the same range. With our quality cut
for UCAC5 proper motion errors, we effectively 
reduced the 
limiting UCAC5 magnitude of the investigated CPM candidates further.
The M/L subdwarfs have in addition
very large proper motions (between $\approx$600\,mas/yr 
and $\approx$3500\,mas/yr) so that we were not expecting to find such 
objects in our CPM search aiming at relatively small proper motions.
The region between WDs and cool subdwarfs in Fig.~\ref{Fig_optnircmd}
should be empty. Therefore, 
all CPM candidates falling in
this CMD close to the arbitrary drawn dividing line between WDs and 
subdwarfs 
have to be taken with caution independent of their possible
chance alignment that we evaluated using Eq.~\ref{Eq_1}.

\section{UPC parallaxes confirming two WD companions}
\label{Sect_upc}

The primaries of our WD companion candidates '1' and '2',
\object{HD 166435} and \object{TYC 3980-1081-1}, were
only recently added to the 25\,pc sample.
The original Hipparcos \citepads{1997ESASP1200.....E}
parallax of the well-known early-G star \object{HD 166435} 
placed it at about 25.2\,pc, whereas the revised Hipparcos 
\citepads{2007A&A...474..653V} results led to a distance 
of 24.8\,pc, and the TGAS parallax finally fixed it
at 24.58\,pc. The Tycho star \object{TYC 3980-1081-1},
was discovered as a close neighbour of the sun
by \citetads{2014AJ....148..119F}, who estimated a photometric
distance of 5.9\,pc, before a first trigonometric parallax
from URAT data was reported by \citetads{2016AJ....151..160F}
giving a distance of about 6.5\,pc. The full UPC catalogue,
including all significant parallaxes determined from the
complete northern hemisphere URAT data, was only recently made 
available \citepads{2016arXiv160406739F,2016yCat.1333....0F}. 
The TGAS parallax of
\object{TYC 3980-1081-1} corresponds to a distance of
8.29\,pc, still well within the 10\,pc limit. From its
absolute magnitude of $M_J=6.94$ we estimate a spectral 
type of $\approx$M2 for \object{TYC 3980-1081-1} using
the relationship between absolute magnitudes and spectral
types from 
\citetads{2005A&A...442..211S}.

Interestingly, not only \object{HD 166435} 
and \object{TYC 3980-1081-1} can be found in the UPC,
but also their WD companion candidates '1' and '2',
which we selected based on their CPM in TGAS and UCAC5. We 
call them \object{HD 166435 B} and \object{TYC 3980-1081-1 B},
respectively. Their UPC parallaxes, albeit much less accurate than
the TGAS parallaxes, are similarly large (see 
upper part of
Table~\ref{Tab_cpm})
as for the primaries and confirm a common distance,
respectively. This is important in particular for
our second candidate, \object{TYC 3980-1081-1 B}, 
which shows a large difference in the 
DE proper motion component compared to that 
of \object{TYC 3980-1081-1}. However, the close distance
and small separation of this CPM pair lead to a
very large $pmo$ value, and the proper motion difference
is not much larger than this expected effect from orbital
motion. In case of the known wide binary 
\object{Ross 429}/\object{SCR J0753-2524}, 
discovered by \citetads{2005AJ....130.1658S},
the proper motion differences are also relatively
large with respect to the corresponding $pmo$ value
(Table~\ref{Tab_cpm}). Note that with our fourth selection 
criterion described in Sect.~\ref{Sect_cpmcriteria} 
we took also into account
the proper motion errors, which are relatively large for
the UCAC5 proper motion of \object{SCR J0753-2524}.

The UPC proper motions of three CPM companions, 
also given in the upper part of Table~\ref{Tab_cpm},
are less accurate than their UCAC5 proper motions. However, for
the two relatively bright WD companions \object{HD 166435 B}
and \object{TYC 3980-1081-1 B} the combined parallax and proper
motion solution in the UPC led to more similar proper motions
compared to their primaries, respectively. For the fainter
known companion \object{WD 0433+270} the UPC errors of its
parallax and proper motion are much larger, and the UCAC5
proper motion is in better agreement with the TGAS proper
motion of the primary than the UPC proper motion.

   \begin{figure}
   \centering
   \includegraphics[width=\hsize]{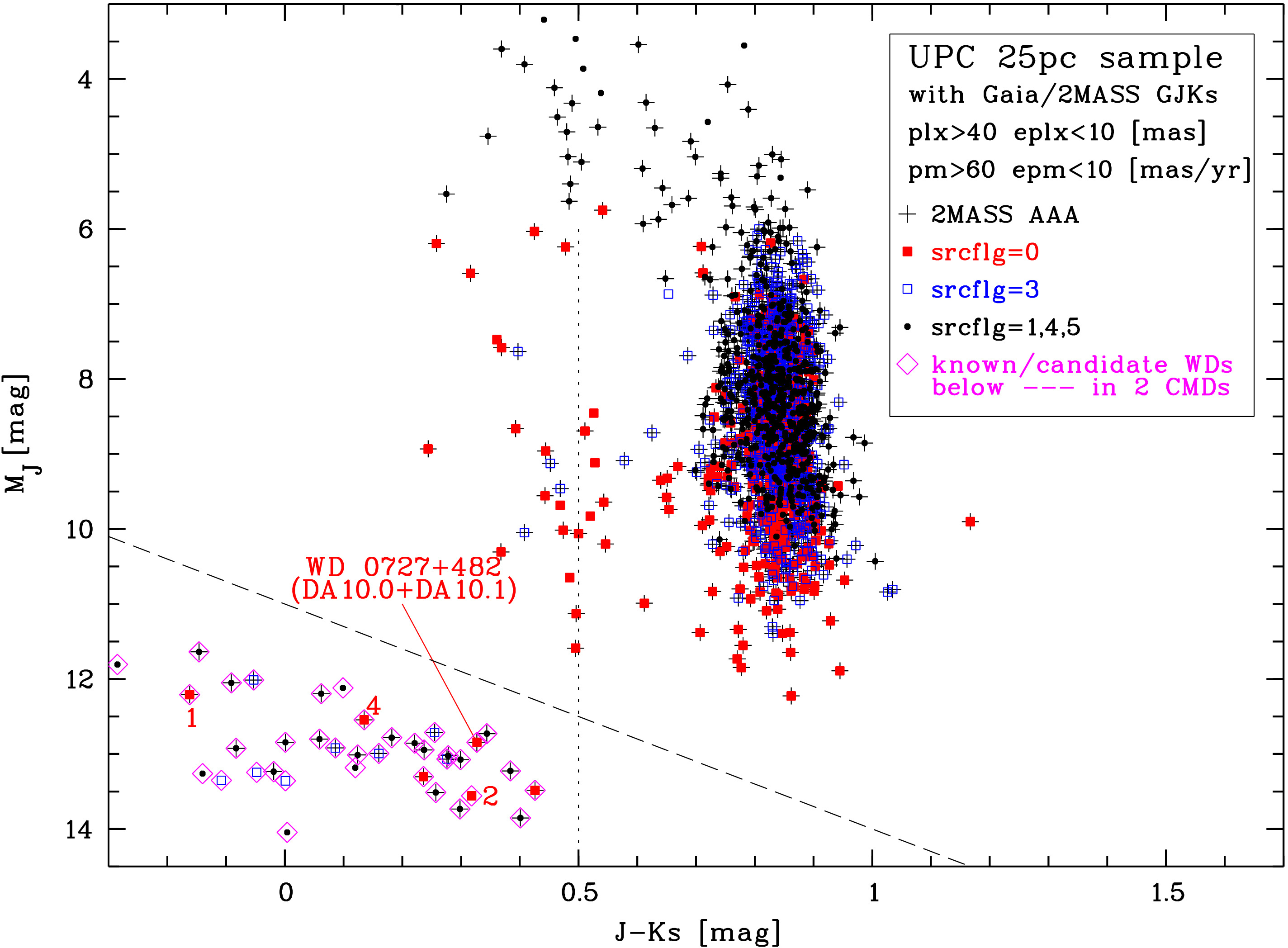}
      \caption{Near-infrared (2MASS) CMD for our selected UPC 25\,pc 
               sample. Absolute magnitudes are based on UPC parallaxes.
               The start and end values of both axes are the same as 
               in Fig.~\ref{Fig_nircmd}, the dashed line shows our previously
               chosen limit for the WD region in this CMD 
               (Eq.~\ref{Eq_3}), 
               the dotted line marks again the
               WD/subdwarf dividing line of \citetads{2017AJ....154...32S}.
               Stars with photometric quality 'AAA' in 2MASS are plotted as
               plus signs. Overplotted as red filled squares, blue open 
               squares and black dots are stars
               without previous parallax measurements (UPC $srcflg=0$),
               with already published data in \citetads{2016AJ....151..160F} 
               ($srcflg=3$), and in the literature ($srcflg=1,4,5$), 
               respectively. 
               All WD candidates, defined by us as falling below the
               dashed lines in this figure and in Fig.~\ref{Fig_optnirupc},
               are shown as open lozenges. The new WD candidates '1' and '2'
               (already found in our TGAS CPM search in UCAC5), and '4'
               are marked. A close binary cool WD is also labelled.}
         \label{Fig_nirupc}
   \end{figure}

   \begin{figure}
   \centering
   \includegraphics[width=\hsize]{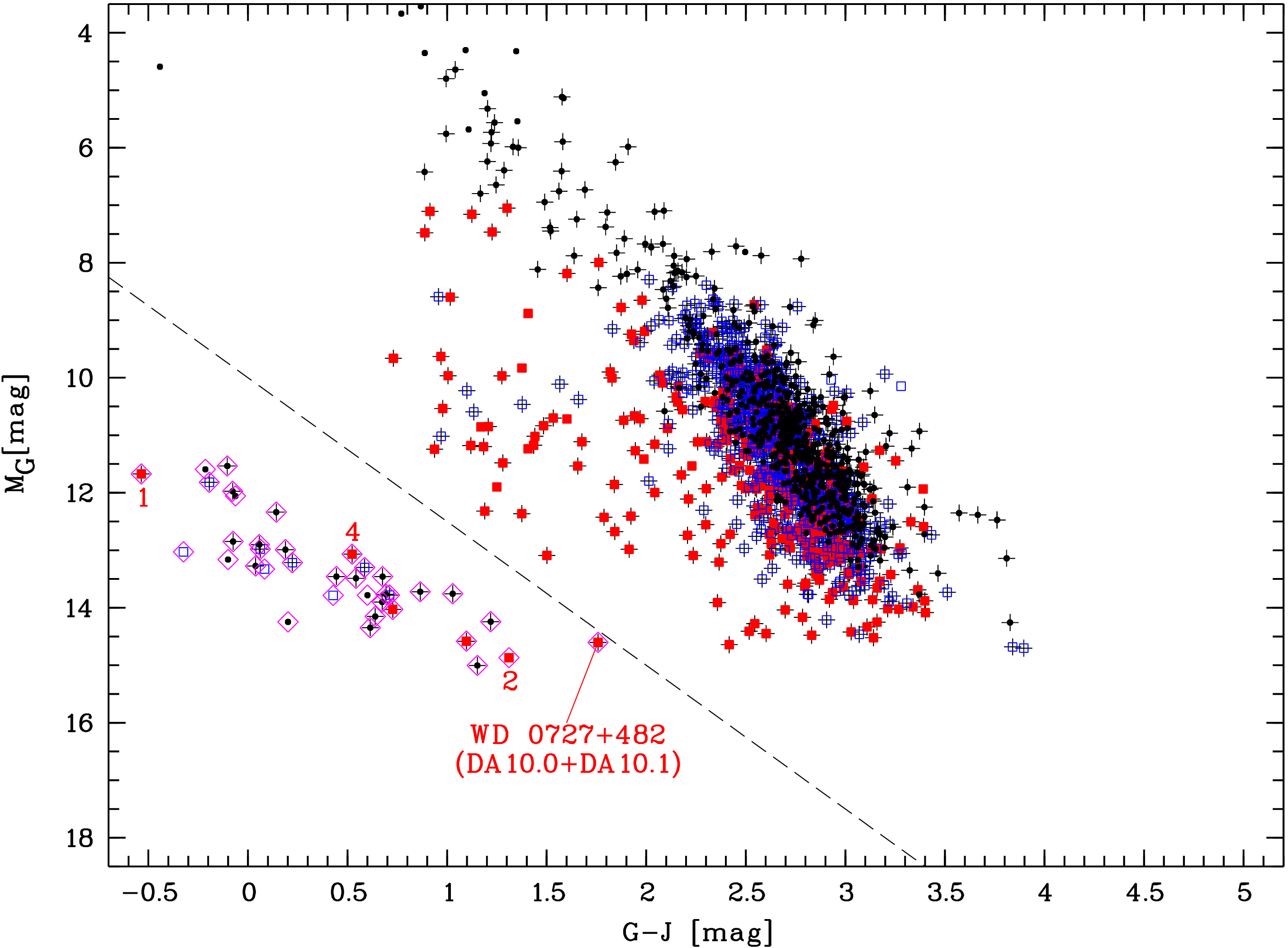}
      \caption{Optical (Gaia) to near-infrared (2MASS) CMD 
               for our selected UPC 25\,pc 
               sample. As in Fig.~\ref{Fig_nirupc}, absolute magnitudes 
               are based on UPC parallaxes. 
               The start and end values of both axes are the same as 
               in Fig.~\ref{Fig_optnircmd}, the dashed line shows again our
               limit for the WD region in this CMD 
               (Eq.~\ref{Eq_4}). 
               Symbols are the same as in Fig.~\ref{Fig_nirupc}.}
         \label{Fig_optnirupc}
   \end{figure}

\section{Additional WD candidates from the UPC}
\label{Sect_addupc}
\smallskip
\citetads{2016AJ....151..160F} have already studied stars
falling in a 25\,pc sample according to significant
parallax measurements in the UPC. They mentioned many new nearby
stars with small proper motions ($<$200\,mas/yr) but did not
pay special attention to potential WDs among their new neighbour
candidates. In their notes on individual systems they discussed
the nearest new discoveries and found many of them suspicious
with respect to blended or elongated images and/or large scatter
in the combined fit for parallax and proper motion. These
suspicious nearest candidates have parallax and proper motion
errors of typically 10-25\,mas and 10-20\,mas/yr, respectively.
From $\approx$112,000 stars in the full UPC  available at the
CDS \citepads{2016yCat.1333....0F}, $\approx$5,200 have parallaxes
$>$40\,mas, and for $\approx$3,700 of them the parallax errors are
smaller than 10\,mas. As the UPC covers only about half of the sky,
this number appears very high in comparison with the already
mentioned RECONS all-sky 25\,pc census, and the new UPC neighbours
need to be taken with caution.

To evaluate potential WDs in near-infrared and optical to
near-inrared CMDs, similar to those shown in Figs~\ref{Fig_nircmd}
and \ref{Fig_optnircmd}, we cross-matched the UPC with 2MASS and Gaia
DR1. We found $\approx$3,700 UPC stars within 25\,pc that have
both 2MASS and Gaia counterparts.
As expected from the two
different parts of the UPC \citepads[stars with or without
previously known parallaxes;
see][]{2016arXiv160406739F,2016yCat.1333....0F} the error distribution
for the UPC parallaxes rises to a peak at 10\,mas before it
abruptly turns down to a long tail reaching to about 60\,mas.
The proper motion
errors show a more symmetric distribution with a maximum at
about 4\,mas/yr, a smooth decrease to 10\,mas/yr, and also
a long tail continuing to about 40\,mas/yr.

For our final UPC 25\,pc sample, consisting of $\approx$1,600
stars,
we applied the following quality criteria
concerning the parallax and proper motion errors, $eplx$ 
and $epm$, respectively, and also used a lower limit
for the total proper motion:

\begin{itemize}
\item $eplx_{UPC}<$10\,mas,
\item $epm_{UPC}<$10\,mas/yr (in RA and DE components),
\item UPC total proper motion $>$60\,mas/yr.
\end{itemize}

The resulting CMDs are shown in Figs.~\ref{Fig_nirupc}
and \ref{Fig_optnirupc}. As in Figs.~\ref{Fig_nircmd} and
\ref{Fig_optnircmd}, we plot stars with reliable 2MASS
photometry as plus signs, but overplot all stars 
independent of their 2MASS quality with different symbols
corresponding to their source flags ($srcflg$) given in the UPC.
Stars with $srcflg=0$, without previously published parallaxes
in SIMBAD or other catalogues, are plotted as filled red squares
(except for \object{GJ 4165} lacking a $G$ magnitude)
and are listed in the lower part of Table~\ref{Tab_cpm} if
not already listed in the upper part. Because of the deeper 
limiting magnitude of URAT compared to UCAC5, the MS is now 
mainly populated with M dwarfs that were lacking in TGAS 
and TGAS/UCAC5 CPM data. With $M_J$=6-11, these are M0-M7
dwarfs according to \citetads{2005A&A...442..211S}. 
As expected from the 
lower accuracy of the UPC parallaxes in comparison to 
TGAS, the MS and the WD sequence in Figs.~\ref{Fig_nirupc} 
and \ref{Fig_optnirupc} do not look as narrow as in 
Figs.~\ref{Fig_nircmd} and \ref{Fig_optnircmd}. In particular,
many of the new UPC neighbours ($srcflg=0$ and $srcflg=3$),
are located left of the MS, where only few subdwarfs are
expected. However,
all our WD candidates defined 
by Eqs.~\ref{Eq_3} and \ref{Eq_4}
are well separated from the cloud of suspicious subdwarfs
left of the MS. The only WD located close to the corresponding
dashed line in Fig.~\ref{Fig_optnircmd}, \object{WD 0727+482},
is a close binary, expected to be brighter. 
We also note that none of the 34 known WDs and 3 WD candidates
('1', '2', and '4') in Fig.~\ref{Fig_nirupc} falls right of the 
WD/subdwarf dividing line of \citetads{2017AJ....154...32S}.
The new candidate '4' is a previously 
anonymous field star, \object{UPC 72924}.

When we tried to allow for larger parallax and proper motion
errors in the UPC 25\,pc sample, e.g. 12\,mas and 12\,mas/yr, 
or 15\,mas and 15\,mas/yr, respectively,
a few additional known WDs were recovered,
They were all falling in the WD regions below the dashed lines in both
CMDs, whereas new WD candidates were not found. On the other hand, a 
lower proper motion limit of 40\,mas/yr led to many more suspicious 
objects falling closer to the dashed lines in both CMDs but only
one more WD candidate that we nevertheless considered as unreliable
because of its relatively large image elongation and large scatter
of the post-fit residuals given in the UPC.

Note that, unless we did not search for CPM candidates in
the UPC alone or with respect to TGAS, one of the known WDs with
a UPC parallax in Table~\ref{Tab_cpm}, \object{WD 0727+482}, being 
itself a close binary, is a member of a wide multiple system. 
It includes \object{GJ 275.2 A} that has a similarly large UPC
parallax and proper motion of 101.2$\pm$4.8\,mas 
and $(-192.5\pm4.4, -1268.4\pm4.1)$\,mas/yr, respectively.

\section{One spectroscopically confirmed WD companion}
\label{Sect_specwd}

Our WD candidate \object{HD 166435 B} was already included in 
our earlier large programme concentrating on the spectroscopic 
classification of nearby cool WD and subdwarf candidates. Within 
this programme it was one of the targets with the smallest proper
motion, at that time selected from an earlier UCAC version. 
Large numbers of comparison objects with known spectral
types were also observed. The low-resolution spectroscopic 
observations were carried out in 2008 (August 8/9) using the 
focal reducer and faint object spectrograph CAFOS mounted 
at the 2.2m telescope
at Calar Alto, Spain. We used the grism B~200 giving a wavelength 
coverage from about 3500\,{\AA} to 7000\,{\AA} and a 
dispersion of 4.7\,{\AA} per pixel. All spectra were reduced 
with standard routines from the ESO MIDAS data reduction package. 

\begin{figure}
   \centering
   \includegraphics[width=\hsize]{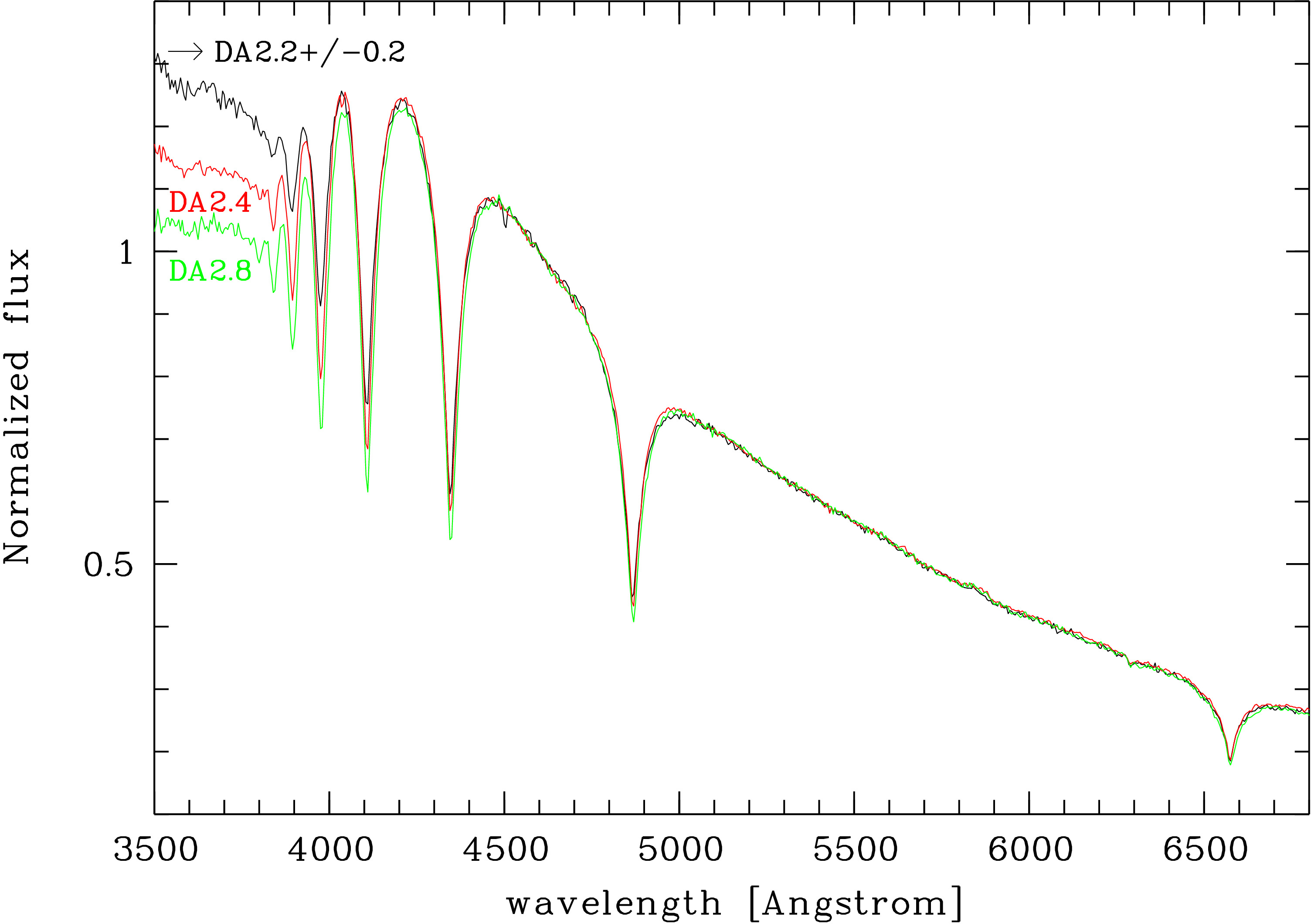}
      \caption{Low resolution Calar Alto 2.2m/CAFOS spectrum of
               candidate '1' (= \object{HD 166435 B}), marked
               on Fig.~\ref{Fig_nircmd} and Fig.~\ref{Fig_optnircmd},
               overplotted by the spectra of two known WDs, 
               \object{WD 2032+248} (DA2.4; red line) and 
               \object{WD 2149+021} (DA2.8; green line) 
               \citepads{2011ApJ...743..138G},
               observed with the same instrument setup. 
               All spectra are normalised at 4600\,{\AA}.
              }
         \label{Fig_wdspec}
   \end{figure}

Fig.~\ref{Fig_wdspec} shows the normalised spectrum of 
\object{HD 166435 B} together with those of two comparison objects
observed during the same observing run. In this comparison
\object{HD 166435 B} appears very similar to the DA2.4 and DA2.8
white dwarfs and could be classified as DA2.0 from its bluer
continuum. The very blue continuum of \object{HD 166435 B}
is consistent with its magnitudes 
and the resulting negative $FUV-NUV$ colour index 
measured by the Galaxy Evolution 
Explorer \citepads[GALEX;][]{2011Ap&SS.335..161B}, $FUV=12.068\pm0.003$
and $NUV=12.735\pm0.004$, compared to $FUV=12.604\pm0.007$
and $NUV=12.688\pm0.005$ 
and a zero color index 
for \object{WD 2149+021} (\object{WD 2032+248} 
was not measured by GALEX).
The distances of \object{HD 166435 B} and \object{WD 2149+021}
are very similar with $\approx$24.6\,pc (TGAS) and
$\approx$24.5\,pc \citepads{2015ApJS..219...19L}, respectively.

We have also measured the equivalent widths of the
H$\beta$ and H$\alpha$ lines, relative to their well-defined 
continuum in the three spectra of Fig.~\ref{Fig_wdspec}. Whereas our
results for H$\alpha$ support a slightly earlier classification
of \object{HD 166435 B} (DA1.9) compared to \object{WD 2032+248}
(DA2.1) and \object{WD 2149+021} (DA2.5), the equivalent 
widths of H$\beta$, measured with higher signal-to-noise, 
hint at DA2.4, DA2.2, and DA2.7, respectively. We assign a
preliminary spectral type of DA2.2$\pm$0.2 to \object{HD 166435 B}.

For the other two WD 
companion 
candidates, \object{TYC 3980-1081-1 B}
and \object{HD 35650 B}, we only assume cool DA types 
from their location in the optical to near-infrared 
CMD (Fig.~\ref{Fig_optnircmd}) in comparison to the known
WDs \object{GJ 171.2 B} and \object{SCR J0753-2524} 
(Table~\ref{Tab_cpm}). 
For the WD candidate \object{UPC 72924} we assume 
a spectral type of $\approx$DA7 from comparing its $G-J$
colour with those of known WDs found in our study of the
UPC 25\,pc sample.


\section{Conclusions}
\label{Sect_concl}

   \begin{enumerate}
      \item Compared to only four known WDs in TGAS and three
            known WD companions in UCAC5, our three new WD companion
            candidates represent an increase of $\approx$43\%.
      \item Two of our new WD companion candidates are confirmed
            by parallax measurements in the UPC. One, \object{HD 166435 B}, 
            orbits a G-type star at a distance of $\approx$24.6\,pc with a 
            projected physical separation of $\approx$700\,AU,
            the other, \object{TYC 3980-1081-1 B}, an early-M dwarf 
            at only $\approx$8.3\,pc distance with a projected physical 
            separation of $\approx$120\,AU, respectively.
            The third WD companion candidate, \object{HD 35650 B},
            has a very large projected physical separation from its primary
            of $\approx$8100\,AU, similar to that of the known wide 
            companion \object{SCR J0753-2524}. 
            However, we can not exclude the possibility
            that \object{HD 35650 B} 
            is only a chance alignment and represents in fact 
            a distant halo star.
      \item Our additional search for nearby WDs using parallaxes
            from the UPC confirmed our WD selection criteria in two CMDs
            already applied for our CPM candidates and led to one more 
            WD candidate, \object{UPC 72924}, at $\approx$25\,pc.
      \item Considering the estimated effect $pmo$ of orbital motion on
            the proper motion difference in CPM pairs, we
            were able to recover the faint components of known very wide 
            CPM systems (\object{GJ 282AB}/\object{GJ 282C} 
            and \object{Ross 429}/\object{SCR J0753-2524}). On the other
            hand, this allowed us to discover \object{TYC 3980-1081-1 B}
            as a cool WD companion candidate with relatively small
            angular separation from a very nearby star, where the $pmo$
            effect is very large.
      \item \object{HD 166435 B} was spectroscopically confirmed as a 
            relatively hot WD, as already suspected from its $G-J=-0.54$
            and large negative GALEX colour index $FUV-NUV\approx-0.67$, 
            whereas \object{TYC 3980-1081-1 B} is with $G-J=+1.31$ a cool 
            WD candidate 
            (possibly $\approx$DA10). 
            Compared to the 
            25\,pc WD sample of \citetads{2016MNRAS.462.2295H}, 
            \object{HD 166435 B} ranges as $\approx$DA2.2 among the hottest
            nearby DA WDs and is an important addition of a 'Sirius-like' 
            system (a WD with a luminous primary of spectral type K or 
            earlier) in the volume between 20\,pc and 25\,pc.
      \item The discovery of \object{TYC 3980-1081-1 B}, if confirmed by 
            follow-up spectroscopy, shows that previous assumptions on the 
            completeness of the WD sample within 13\,pc may be not correct.
            This concerns in particular cool WDs with small proper motions.
      \item The identification of relatively close companions
            to very bright stars, such as \object{HD 166435 B} and
            \object{TYC 3980-1081-1 B} has always been a challenge
            with photographic all-sky surveys. We took advantage of the
            better resolution of CCD images in UCAC5 and Gaia. Gaia's
            next data releases will certainly help identifying more close
            WD companions of nearby stars. On the other hand, 
            \citetads{2015ApJS..219...19L} found eight {\it single} WDs
            as probable new members of the 25\,pc sample in their
            photographic (SUPERBLINK) survey of the Northern sky
            with a lower proper motion limit of 40\,mas/yr.
            A similar number of missing nearby WDs
            can be expected in the Southern sky,
            and additional WDs with even smaller proper motions 
            and trigonometric distances $<$25\,pc may still be
            discovered in forthcoming Gaia all-sky catalogues.
   \end{enumerate}

\begin{acknowledgements}
We have extensively used the SIMBAD database and the VizieR catalogue
access tool operated at the CDS/Strasbourg.
The spectroscopic confirmation of one of our WD candidates was part of
a larger observing campaign carried out with the
2.2\,m telescope of the German-Spanish Astronomical Centre at Calar Alto, 
Spain.
Part of the observations were carried out in service mode. We would
like to thank the Calar Alto staff for their
kind support and for their help with the observations.
We thank Axel Schwope for providing equivalent widths of Balmer lines
in DA WDs.
It is a pleasure to thank Benjamin Braun and Peter Grodzewitz for their
help with a first look at the TGAS and TGAS/UCAC5 nearby subsamples
during their internships at AIP in February and May 2017, respectively.
We would like to thank the referee for many valuable comments and 
suggestions that helped us improve the paper.
\end{acknowledgements}

%
%

\bibliographystyle{aa}
\bibliography{tgucwd}

\begin{thebibliography}{44}
\expandafter\ifx\csname natexlab\endcsname\relax\def\natexlab#1{#1}\fi

\bibitem[{{Andrews} {et~al.}(2017){Andrews}, {Chanam{\'e}}, \&
  {Ag{\"u}eros}}]{2017MNRAS.472..675A}
{Andrews}, J.~J., {Chanam{\'e}}, J., \& {Ag{\"u}eros}, M.~A. 2017, \mnras, 472,
  675

\bibitem[{{Bianchi} {et~al.}(2011){Bianchi}, {Herald}, {Efremova}, {Girardi},
  {Zabot}, {Marigo}, {Conti}, \& {Shiao}}]{2011Ap&SS.335..161B}
{Bianchi}, L., {Herald}, J., {Efremova}, B., {et~al.} 2011, \apss, 335, 161

\bibitem[{{Dupuy} \& {Liu}(2012)}]{2012ApJS..201...19D}
{Dupuy}, T.~J. \& {Liu}, M.~C. 2012, \apjs, 201, 19

\bibitem[{{ESA}(1997)}]{1997ESASP1200.....E}
{ESA}, ed. 1997, ESA Special Publication, Vol. 1200, {The HIPPARCOS and TYCHO
  catalogues. Astrometric and photometric star catalogues derived from the ESA
  HIPPARCOS Space Astrometry Mission}

\bibitem[{{Finch} \& {Zacharias}(2016{\natexlab{a}})}]{2016arXiv160406739F}
{Finch}, C. \& {Zacharias}, N. 2016{\natexlab{a}}, ArXiv e-prints
  [\eprint[arXiv]{1604.06739}]

\bibitem[{{Finch} \& {Zacharias}(2016{\natexlab{b}})}]{2016AJ....151..160F}
{Finch}, C.~T. \& {Zacharias}, N. 2016{\natexlab{b}}, \aj, 151, 160

\bibitem[{{Finch} \& {Zacharias}(2016{\natexlab{c}})}]{2016yCat.1333....0F}
{Finch}, C.~T. \& {Zacharias}, N. 2016{\natexlab{c}}, VizieR Online Data
  Catalog, 1333

\bibitem[{{Finch} {et~al.}(2014){Finch}, {Zacharias}, {Subasavage}, {Henry}, \&
  {Riedel}}]{2014AJ....148..119F}
{Finch}, C.~T., {Zacharias}, N., {Subasavage}, J.~P., {Henry}, T.~J., \&
  {Riedel}, A.~R. 2014, \aj, 148, 119

\bibitem[{{Gaia Collaboration} {et~al.}(2016{\natexlab{a}}){Gaia
  Collaboration}, {Brown}, {Vallenari}, {Prusti}, {de Bruijne}, {Mignard},
  {Drimmel}, {Babusiaux}, {Bailer-Jones}, {Bastian}, \&
  et~al.}]{2016A&A...595A...2G}
{Gaia Collaboration}, {Brown}, A.~G.~A., {Vallenari}, A., {et~al.}
  2016{\natexlab{a}}, \aap, 595, A2

\bibitem[{{Gaia Collaboration} {et~al.}(2016{\natexlab{b}}){Gaia
  Collaboration}, {Prusti}, {de Bruijne}, {Brown}, {Vallenari}, {Babusiaux},
  {Bailer-Jones}, {Bastian}, {Biermann}, {Evans}, \&
  et~al.}]{2016A&A...595A...1G}
{Gaia Collaboration}, {Prusti}, T., {de Bruijne}, J.~H.~J., {et~al.}
  2016{\natexlab{b}}, \aap, 595, A1

\bibitem[{{Gaia Collaboration} {et~al.}(2017){Gaia Collaboration}, {van
  Leeuwen}, {Vallenari}, {Jordi}, {Lindegren}, {Bastian}, {Prusti}, {de
  Bruijne}, {Brown}, {Babusiaux}, \& et~al.}]{2017A&A...601A..19G}
{Gaia Collaboration}, {van Leeuwen}, F., {Vallenari}, A., {et~al.} 2017, \aap,
  601, A19

\bibitem[{{Gaidos} {et~al.}(2014){Gaidos}, {Mann}, {L{\'e}pine}, {Buccino},
  {James}, {Ansdell}, {Petrucci}, {Mauas}, \& {Hilton}}]{2014MNRAS.443.2561G}
{Gaidos}, E., {Mann}, A.~W., {L{\'e}pine}, S., {et~al.} 2014, \mnras, 443, 2561

\bibitem[{{Gianninas} {et~al.}(2011){Gianninas}, {Bergeron}, \&
  {Ruiz}}]{2011ApJ...743..138G}
{Gianninas}, A., {Bergeron}, P., \& {Ruiz}, M.~T. 2011, \apj, 743, 138

\bibitem[{{Gliese} \& {Jahrei{\ss}}(1995)}]{1995yCat.5070....0G}
{Gliese}, W. \& {Jahrei{\ss}}, H. 1995, VizieR Online Data Catalog, 5070

\bibitem[{{Gray} {et~al.}(2006){Gray}, {Corbally}, {Garrison}, {McFadden},
  {Bubar}, {McGahee}, {O'Donoghue}, \& {Knox}}]{2006AJ....132..161G}
{Gray}, R.~O., {Corbally}, C.~J., {Garrison}, R.~F., {et~al.} 2006, \aj, 132,
  161

\bibitem[{{Gray} {et~al.}(2003){Gray}, {Corbally}, {Garrison}, {McFadden}, \&
  {Robinson}}]{2003AJ....126.2048G}
{Gray}, R.~O., {Corbally}, C.~J., {Garrison}, R.~F., {McFadden}, M.~T., \&
  {Robinson}, P.~E. 2003, \aj, 126, 2048

\bibitem[{{Henry} \& {Jao}(2015)}]{2015IAUGA..2253773H}
{Henry}, T.~J. \& {Jao}, W.-C. 2015, IAU General Assembly, 22, 2253773

\bibitem[{{Holberg} {et~al.}(2016){Holberg}, {Oswalt}, {Sion}, \&
  {McCook}}]{2016MNRAS.462.2295H}
{Holberg}, J.~B., {Oswalt}, T.~D., {Sion}, E.~M., \& {McCook}, G.~P. 2016,
  \mnras, 462, 2295

\bibitem[{{Kervella} {et~al.}(2017){Kervella}, {Th{\'e}venin}, \&
  {Lovis}}]{2017A&A...598L...7K}
{Kervella}, P., {Th{\'e}venin}, F., \& {Lovis}, C. 2017, \aap, 598, L7

\bibitem[{{L{\'e}pine}(2011)}]{2011ASPC..448.1375L}
{L{\'e}pine}, S. 2011, in Astronomical Society of the Pacific Conference
  Series, Vol. 448, 16th Cambridge Workshop on Cool Stars, Stellar Systems, and
  the Sun, ed. C.~{Johns-Krull}, M.~K. {Browning}, \& A.~A. {West}, 1375

\bibitem[{{L{\'e}pine}(2017)}]{2017AAS...22915601L}
{L{\'e}pine}, S. 2017, in American Astronomical Society Meeting Abstracts, Vol.
  229, American Astronomical Society Meeting Abstracts, 156.01

\bibitem[{{L{\'e}pine} \& {Bongiorno}(2007)}]{2007AJ....133..889L}
{L{\'e}pine}, S. \& {Bongiorno}, B. 2007, \aj, 133, 889

\bibitem[{{L{\'e}pine} \& {Shara}(2005)}]{2005AJ....129.1483L}
{L{\'e}pine}, S. \& {Shara}, M.~M. 2005, \aj, 129, 1483

\bibitem[{{Li} {et~al.}(2014){Li}, {Marshall}, {L{\'e}pine}, {Williams}, \&
  {Chavez}}]{2014AJ....148...60L}
{Li}, T., {Marshall}, J.~L., {L{\'e}pine}, S., {Williams}, P., \& {Chavez}, J.
  2014, \aj, 148, 60

\bibitem[{{Limoges} {et~al.}(2015){Limoges}, {Bergeron}, \&
  {L{\'e}pine}}]{2015ApJS..219...19L}
{Limoges}, M.-M., {Bergeron}, P., \& {L{\'e}pine}, S. 2015, \apjs, 219, 19

\bibitem[{{Lindegren} {et~al.}(2016){Lindegren}, {Lammers}, {Bastian},
  {Hern{\'a}ndez}, {Klioner}, {Hobbs}, {Bombrun}, {Michalik}, {Ramos-Lerate},
  {Butkevich}, {Comoretto}, {Joliet}, {Holl}, {Hutton}, {Parsons},
  {Steidelm{\"u}ller}, {Abbas}, {Altmann}, {Andrei}, {Anton}, {Bach},
  {Barache}, {Becciani}, {Berthier}, {Bianchi}, {Biermann}, {Bouquillon},
  {Bourda}, {Br{\"u}semeister}, {Bucciarelli}, {Busonero}, {Carlucci},
  {Casta{\~n}eda}, {Charlot}, {Clotet}, {Crosta}, {Davidson}, {de Felice},
  {Drimmel}, {Fabricius}, {Fienga}, {Figueras}, {Fraile}, {Gai}, {Garralda},
  {Geyer}, {Gonz{\'a}lez-Vidal}, {Guerra}, {Hambly}, {Hauser}, {Jordan},
  {Lattanzi}, {Lenhardt}, {Liao}, {L{\"o}ffler}, {McMillan}, {Mignard}, {Mora},
  {Morbidelli}, {Portell}, {Riva}, {Sarasso}, {Serraller}, {Siddiqui}, {Smart},
  {Spagna}, {Stampa}, {Steele}, {Taris}, {Torra}, {van Reeven}, {Vecchiato},
  {Zschocke}, {de Bruijne}, {Gracia}, {Raison}, {Lister}, {Marchant},
  {Messineo}, {Soffel}, {Osorio}, {de Torres}, \&
  {O'Mullane}}]{2016A&A...595A...4L}
{Lindegren}, L., {Lammers}, U., {Bastian}, U., {et~al.} 2016, \aap, 595, A4

\bibitem[{{Luyten}(1995)}]{1995yCat.1098....0L}
{Luyten}, W.~J. 1995, VizieR Online Data Catalog, 1098

\bibitem[{{Luyten}(1997)}]{1997yCat.1130....0L}
{Luyten}, W.~J. 1997, VizieR Online Data Catalog, 1130

\bibitem[{{Luyten}(1998)}]{1998yCat.1087....0L}
{Luyten}, W.~J. 1998, VizieR Online Data Catalog, 1087

\bibitem[{{Nelan} {et~al.}(2015){Nelan}, {Bond}, \&
  {Schaefer}}]{2015ASPC..493..501N}
{Nelan}, E.~P., {Bond}, H.~E., \& {Schaefer}, G. 2015, in Astronomical Society
  of the Pacific Conference Series, Vol. 493, 19th European Workshop on White
  Dwarfs, ed. P.~{Dufour}, P.~{Bergeron}, \& G.~{Fontaine}, 501

\bibitem[{{Oh} {et~al.}(2017){Oh}, {Price-Whelan}, {Hogg}, {Morton}, \&
  {Spergel}}]{2017AJ....153..257O}
{Oh}, S., {Price-Whelan}, A.~M., {Hogg}, D.~W., {Morton}, T.~D., \& {Spergel},
  D.~N. 2017, \aj, 153, 257

\bibitem[{{Poveda} {et~al.}(2009){Poveda}, {Allen}, {Costero},
  {Echevarr{\'{\i}}a}, \& {Hern{\'a}ndez-Alc{\'a}ntara}}]{2009ApJ...706..343P}
{Poveda}, A., {Allen}, C., {Costero}, R., {Echevarr{\'{\i}}a}, J., \&
  {Hern{\'a}ndez-Alc{\'a}ntara}, A. 2009, \apj, 706, 343

\bibitem[{{Scholz} {et~al.}(2005){Scholz}, {Meusinger}, \&
  {Jahrei{\ss}}}]{2005A&A...442..211S}
{Scholz}, R.-D., {Meusinger}, H., \& {Jahrei{\ss}}, H. 2005, \aap, 442, 211

\bibitem[{{Skrutskie} {et~al.}(2006){Skrutskie}, {Cutri}, {Stiening},
  {Weinberg}, {Schneider}, {Carpenter}, {Beichman}, {Capps}, {Chester},
  {Elias}, {Huchra}, {Liebert}, {Lonsdale}, {Monet}, {Price}, {Seitzer},
  {Jarrett}, {Kirkpatrick}, {Gizis}, {Howard}, {Evans}, {Fowler}, {Fullmer},
  {Hurt}, {Light}, {Kopan}, {Marsh}, {McCallon}, {Tam}, {Van Dyk}, \&
  {Wheelock}}]{2006AJ....131.1163S}
{Skrutskie}, M.~F., {Cutri}, R.~M., {Stiening}, R., {et~al.} 2006, \aj, 131,
  1163

\bibitem[{{Smart} {et~al.}(2017){Smart}, {Marocco}, {Caballero}, {Jones},
  {Barrado}, {Beam{\'{\i}}n}, {Pinfield}, \& {Sarro}}]{2017MNRAS.469..401S}
{Smart}, R.~L., {Marocco}, F., {Caballero}, J.~A., {et~al.} 2017, \mnras, 469,
  401

\bibitem[{{Subasavage} {et~al.}(2005){Subasavage}, {Henry}, {Hambly}, {Brown},
  {Jao}, \& {Finch}}]{2005AJ....130.1658S}
{Subasavage}, J.~P., {Henry}, T.~J., {Hambly}, N.~C., {et~al.} 2005, \aj, 130,
  1658

\bibitem[{{Subasavage} {et~al.}(2009){Subasavage}, {Jao}, {Henry}, {Bergeron},
  {Dufour}, {Ianna}, {Costa}, \& {M{\'e}ndez}}]{2009AJ....137.4547S}
{Subasavage}, J.~P., {Jao}, W.-C., {Henry}, T.~J., {et~al.} 2009, \aj, 137,
  4547

\bibitem[{{Subasavage} {et~al.}(2017){Subasavage}, {Jao}, {Henry}, {Harris},
  {Dahn}, {Bergeron}, {Dufour}, {Dunlap}, {Barlow}, {Ianna}, {L{\'e}pine}, \&
  {Margheim}}]{2017AJ....154...32S}
{Subasavage}, J.~P., {Jao}, W.-C., {Henry}, T.~J., {et~al.} 2017, \aj, 154, 32

\bibitem[{{Taylor}(2005)}]{2005ASPC..347...29T}
{Taylor}, M.~B. 2005, in Astronomical Society of the Pacific Conference Series,
  Vol. 347, Astronomical Data Analysis Software and Systems XIV, ed.
  P.~{Shopbell}, M.~{Britton}, \& R.~{Ebert}, 29

\bibitem[{{Torres} {et~al.}(2006){Torres}, {Quast}, {da Silva}, {de La Reza},
  {Melo}, \& {Sterzik}}]{2006A&A...460..695T}
{Torres}, C.~A.~O., {Quast}, G.~R., {da Silva}, L., {et~al.} 2006, \aap, 460,
  695

\bibitem[{{Tremblay} {et~al.}(2017){Tremblay}, {Gentile-Fusillo}, {Raddi},
  {Jordan}, {Besson}, {G{\"a}nsicke}, {Parsons}, {Koester}, {Marsh}, {Bohlin},
  {Kalirai}, \& {Deustua}}]{2017MNRAS.465.2849T}
{Tremblay}, P.-E., {Gentile-Fusillo}, N., {Raddi}, R., {et~al.} 2017, \mnras,
  465, 2849

\bibitem[{{van Leeuwen}(2007)}]{2007A&A...474..653V}
{van Leeuwen}, F. 2007, \aap, 474, 653

\bibitem[{{Zacharias} {et~al.}(2017){Zacharias}, {Finch}, \&
  {Frouard}}]{2017AJ....153..166Z}
{Zacharias}, N., {Finch}, C., \& {Frouard}, J. 2017, \aj, 153, 166

\bibitem[{{Zacharias} {et~al.}(2015){Zacharias}, {Finch}, {Subasavage},
  {Bredthauer}, {Crockett}, {Divittorio}, {Ferguson}, {Harris}, {Harris},
  {Henden}, {Kilian}, {Munn}, {Rafferty}, {Rhodes}, {Schultheiss}, {Tilleman},
  \& {Wieder}}]{2015AJ....150..101Z}
{Zacharias}, N., {Finch}, C., {Subasavage}, J., {et~al.} 2015, \aj, 150, 101

\end{thebibliography}
\end{document}